# What drives adsorption of ions on surface of nanodiamonds in aqueous solutions?

Farshad Saberi-Movahed[*] and Donald W Brenner

Department of Materials Science and Engineering, North Carolina State University, Raleigh, NC, USA.

([*]Corresponding author: fsaberi@ncsu.edu)

**ABSTRACT**

It is not yet clear what drives the adsorption of ions on detonation nanodiamonds (DNDs), which plays a critical role on the loading (unloading) of chemotherapeutic drugs on (from) the surface of DNDs in their targeted therapy applications. Furthermore, effects of adsorbed ions on the hydration layers of water around DNDs with different surface chemistries have not been studied yet. Through a series of Molecular Dynamics simulations, we found out that the law of matching water affinity generally explains well the adsorption patterns of ions onto the surface functional groups of DNDs. Depending on whether the water affinity of the ion matches with that of the surface functional group or not, the former predominantly forms either Contact Ion-Pair (CIP) or Solvent-shared Ion-Pair (SIP) with the latter. In this regard, $Na^+$ and $Mg^{2+}$ have the highest tendencies to form, respectively, CIP and SIP associations with $-COO^-$ functional groups. In the extreme case of 84 $-COO^-$ groups on DND–COOH, however, we observed few $Mg^{2+}$–$COO^-$ CIP associations, for which we have proposed a hypothesis based on the entropy gains. Furthermore, $Mg^{2+}$ and to a lesser extent $Ca^{2+}$ in cooperation with $-COO^-$ functional groups on the surface of charged DND–COOH lead to relatively high residence times of water in the first hydration layer of DND. This study also provides a firsthand molecular level insight about the preferential orientation of water in the vicinity of positively charged DND–H, on which prior experimental studies have not yet reached a consensus.

## 1. Introduction

Over the past two decades, the Detonation Nanodiamond (DND) particles have drawn growing attention in science and technology communities, due to their unique properties. Particularly, they have great amenability to targeted surface functionalization or to non-covalently adsorb some adsorbates such as drugs or impurities.[1] This has made them appealing candidates for such applications as anticancer drug carriers[2–5], protein delivery vehicles[6], bioimaging agents[7–9], nanolubricants, sorbents in chromatography[10–14], and lubricant additives[15–17].

In aforementioned applications, DND particles often interact with an aqueous solution of some sort of ions and possibly some other residues such as proteins. Furthermore, DNDs are often commercially available in the form of aqueous solutions, whose colloidal stabilities are of great concerns.[18–20] Therefore, it is imperative to understand interactions between DNDs and their surrounding aqueous solutions. In particular, many studies have found that solutes typically induce perturbations in the hydrogen bond (HB) network of their surrounding water. This leads to the formation of structured hydration shells around solutes, which reportedly can extend up to 1-2 nm from the solute's surfaces.[21–23] Indeed, the structural and dynamic characteristics of hydration shells around solutes have been associated with a whole host of properties in aqueous solutions.[24–29] For instance, Liu *et al.* recently reported that the first and second hydration shells around $TiO_2$ nanoparticles can create entropic barriers for the adsorption of amino acids with acidic residues.[30]

Despite the rich literature available in the field of colloidal solution of DNDs, few studies have been devoted to the hydration shells of DNDs. One of the earliest records can be found in Korobov *et al.*'s study[31]. They reported the formation of a nanophase of water as a 1 nm thick spherical shell around a DND (with 5.2 nm in diameter) in an aqueous gel. Then, they attributed the colloidal stability of this solution to a smaller spherical shell within that nanophase of water, which remained intact during differential scanning calorimetry experiments. In a more recent study, Stehlik *et al.* also carried out some thermal analysis experiments on the hydrogenated and oxidized DNDs and observed different water adsorption patterns on these DNDs.[32] They found higher water content and weaker interactions of water with the former compared with the latter. Furthermore, long-range disruptions in the HB network of water around positively charged hydrogenated DNDs have been reported.[33,34] In particular, Petit *et al.* predicted that water molecules orient their dangling OH toward the surface of the aforementioned DNDs, which results in weakening of HBs in the interfacial water with respect to the bulk water.[33] However, Chaux-Jukic *et al.* have recently reported a completely opposite preferential orientation for hydrating water of hydrogenated DNDs.[35]

Despite great efforts mentioned above, a complete picture at the molecular level about the structure of hydration layers of DNDs is still missing. In addition, the effects of salts and their constituent ions on these hydration layers have not yet been studied. The investigation of these effects is crucial from two different, yet pertinent perspectives.

First, ions are indispensable ingredients in most biomedical and sorbent applications of DNDs.[36,37] For instance, a couple of studies have shown that certain salts such as NaCl can give superior drug loading capabilities to DNDs.[2,38] Nonetheless, Zhu *et al.* have reported that excessive adsorbed $Na^+$ ions on DNDs, which carry an anti-cancer drug, can damage cells in a serum-free medium.[39] Indeed, ions' adsorption and retention rates on DNDs' surfaces differ depending on the charge density and the size of ions and also the type of the DND's surface chemistry.[10,40]

Second, ions perturb the structural and dynamic characteristics of the surrounding water, based on which they have been classified into kosmotropes (structure-maker) or chaotropes (structure-breaker).[41,42] The former are ions with high charge densities such as $Mg^{2+}$ or $Ca^{2+}$ that impart an ordered water in their immediate vicinity. In contrast, chaotropes such as $K^+$ or $Cl^-$ weakly bind to water due to their low charge densities, which leads to a disorganized water arrangement around them. These "specific ion effects"[43] can significantly modify the properties of electrolyte solutions.[34,41,44–46] For instance, aqueous solutions containing structure-makers such as $Mg^{2+}$, $Ca^{2+}$ or $SO_4^{2-}$ have substantially lower self-diffusion coefficients and higher viscosities compared with pure water.[47–49]

We therefore set out to study the hydration layers of DNDs in aqueous solutions of different inorganic salts. We focus on the effects of the DND's surface chemistry and the solvated salt on the structure of hydration layers and the adsorption behavior of ions therein. To this end, we have employed Molecular Dynamics (MD) simulations[50], which can benefit us to obtain insights into the hydration layers with atomistic details. In particular, MD simulations enable us to obtain the density of water in hydration layers with respect to that of bulk water, different adsorption modes of ions on the surface of DNDs, the preferential orientations of the interfacial water, and the mean residence time of water in hydration layers.

The rest of this paper is organized as follows. In Section 2, we have presented the system setup, MD simulation details, and the computational tools to study the properties of interest. Then, in Section 3.1, we have investigated two density measures: Perpendicular Number Density (PND), partial Radial Distribution Function (RDF). The PND is used to characterize the structure of hydration layers in the vicinity of DND facets. We use the partial RDFs to mainly study the association behavior of dissolved ions with themselves and also with constituent atoms of DND surface functional groups. In Section 3.2, we have studied the preferential orientation of DNDs' interfacial water using two distributions: 1) the Angular Distribution (AD) as a normalized 1D histogram, 2) the Distance-dependent Angular Distribution (DAD) as a normalized 2D histogram. In Section 3.3, we have investigated the residence time of water in hydration layers of DNDs. It qualitatively gives us a measure of how tightly bound the hydration layers are to the DND's surfaces. In Section 4, we have presented our concluding remarks.

## 2. Methodology

### 2.1. MD simulation setup

Each system in our MD simulations consists of a single DND with a specific surface functionalization and a particular number of charges that is solvated in a cubic box of 0.1M aqueous solution of one of KCl, NaCl, $CaCl_2$, or $MgCl_2$ salts. The whole simulation box is comprised of 53781 water molecules on average. A snapshot of such a system is shown in Figure 1, for the neutral hydrogenated DND that is solvated in NaCl salt solution. The motivation behind choosing the aforementioned salts is twofold. First, their constituent ions are ubiquitous in the human body with vital roles in physiological processes[51,52], which is important for biomedical applications of DNDs. Second, $K^+$ and $Cl^-$ are considered structure-breakers, while other constituent ions of these salts are structure-makers. The specific ion effects of these ions at various solid/water and air/water interfaces with microscopic details are well-documented in the literature.[53–56] However, a study of such effects on hydration layers of DNDs is missing.

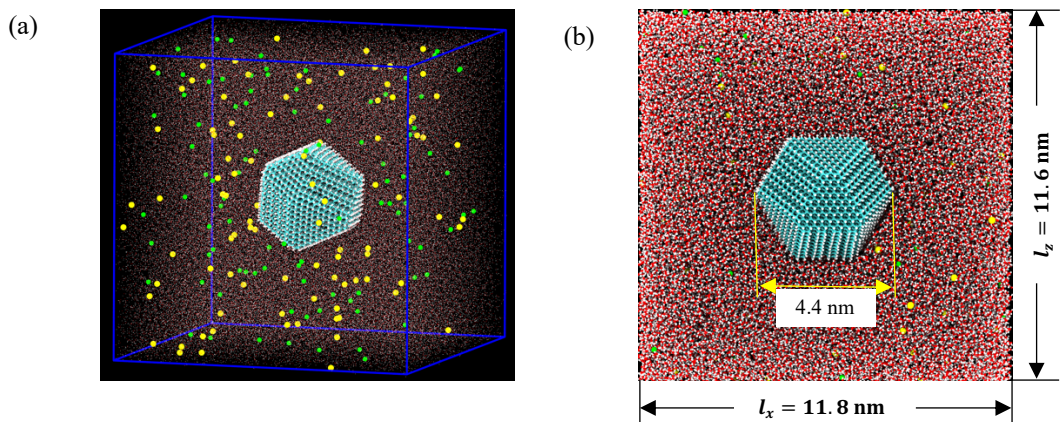

**Figure 1.** (a) Atomistic representation of a typical MD system in our study, which consists of one DND particle (here, neutral hydrogenated DND) solvated in a cubic box of 0.1M aqueous solution of one of KCl, NaCl, $CaCl_2$, or $MgCl_2$ salts (here, NaCl salt), (b) a cut halfway through the water box. Particles with cyan, red, white, yellow, and green colors represent carbon, oxygen, hydrogen, cation and anion atoms, respectively.

The DND in our study has a cuboctahedral shape with 6 and 8 facets with {100} and {111} crystallographic orientations, respectively, where the former are undergone the so-called 2x1 dimerization.[57–60] We have considered four different, commonly used functional groups, which are hydrogen atom (–H), amine group (–$NH_2$), carboxyl group (–COOH), and hydroxyl group (–OH). They are connected with a single covalent bond to carbon atoms on DNDs' surfaces that have one dangling bond. Hereafter, we refer to DNDs with these functional groups as DND–H, DND–$NH_2$, DND–COOH, and DND–OH, respectively. We have created atomic structures of the last three

DNDs by substituting their corresponding polar groups for some H atoms on DND–H. Thus, they still have some hydrogen atoms remaining on their surfaces.

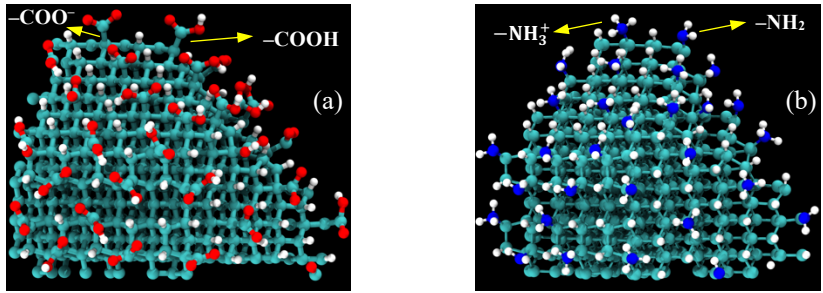

**Figure 2.** (a) Atomistic representation of a portion of charged DND–COOH's surface, (b) same as (a) but for charged DND–NH$_2$. Particles with cyan, red, white, and blue colors represent carbon, oxygen, hydrogen, and nitrogen atoms, respectively.

All DNDs in this study exist as both neutral and charged particles, except for DND–OH that has zero net charge. Charged DNDs have net absolute charges of 28, 56, and 84, with the positive sign for DND–H and DND–NH$_2$ and the negative sign for DND–COOH. Each unit of the absolute charge on DND–NH$_2$ and DND–COOH particles is obtained by exactly substituting one $-NH_3^+$ group and one $-COO^-$ group for, respectively, one $-NH_2$ group and one $-COOH$ group. A portion of charged DND–NH$_2$ and charged DND–COOH are presented in Figure 2. We have assigned positive charges to DND–H using a different procedure to somehow resemble the theory behind the nature of positive charges in hydrogenated DNDs. Indeed, Petit *et. al.* have reported a positive zeta potential (ca. +40 mV) for DND–H in water.[61] They have attributed this observation to hole accumulations on DND's surfaces as a result of electron exchanges at those surfaces with an electrochemical redox couple involving oxygen in water. Thus, we have adopted a procedure developed by Su *et. al.*[62] to assign partial charges to C and H atoms on surfaces of DND–H such that the resulting DND possesses net charges of +28, +56, or +84.

We carried out MD simulations using LAMMPS package[63], and OPLS-AA forcefield[64] to describe both intramolecular and intermolecular interactions for DNDs. The former involve the bond stretching, angle bending, and torsional potentials, while the latter correspond to Coulomb and LJ potentials for non-bonded interactions. We treated water in our MD simulations as an explicit solvent, since we want to study hydration layers with atomistic details. Hence, we employed the SPC/E model[65] to describe the atomic structure of water and also its interactions with other species. Our choice of this model is based on its capability to reproduce the structure and dynamics of bulk water that agree well with experimental measurements[66,67]. Interactions of ions with themselves and also with other atoms were modeled by LJ and Coulomb potentials. We adopted LJ parameters for ions from Ref.[68] for NaCl and KCl salts and from Ref.[69] for MgCl$_2$ and CaCl$_2$ salts.

We treated the long-range electrostatic interactions using the Particle-Particle Particle-Mesh (PPPM) method[70] with the accuracy of 0.3 in the desired relative error in forces. Moreover, we considered an identical cut-off distance for all LJ interactions to be 2.5 times the average LJ diameter of distinct atoms (i.e., $\sigma_{ii}$) in the system. Cross-term LJ parameters, i.e., $\sigma_{ij}$ and $\varepsilon_{ij}$ for $i \neq j$ were calculated using the Lorentz-Berthelot combining rule[71]. That is,

$$\sigma_{ij} = \frac{1}{2}(\sigma_{ii} + \sigma_{jj}) \qquad \text{Eq. 1}$$

$$\varepsilon_{ij} = \sqrt{\varepsilon_{ii}\varepsilon_{jj}} \qquad \text{Eq. 2}$$

We set the MD time step to 1 femtosecond (fs) and chose the "velocity-Verlet" algorithm[72] to integrate the equations of motion. We first let the system equilibrate at 298 K and 1 bar in the NPT ensemble for 2 nanoseconds (ns). The Nose-Hoover[73,74] thermostat and barostat were used to maintain the temperature and pressure of the system at target values with damping factors of 0.1 and 1.0 picosecond (ps) for the former and latter, respectively. The equilibration phase helped relax atomic structures and also facilitated the system to reach its equilibrium density at the target physical conditions. Then, the production phase followed for 2 ns in the NVT ensemble, where the temperature was maintained at 298 K using the Nose-Hoover thermostat.

## 2.2. Computational Tools

We developed some computational tools with the Message Passing Interface (MPI) capability for parallel computing as adds-on to the MDAnalysis package[75,76]. These tools helped us post-process the atomic trajectories in order to characterize the hydration layers of water around different solutes in this study.

### 2.2.1. Radial Distribution Function

We employ the partial Radial Distribution Function (RDF), also known as the partial pair correlation function, to characterize the structural correlations between tagged particle pairs that exist in our atomic systems. In particular, it gives us a quantitative measure of the extent of perturbations that a specific particle can exert in the atomic structure of its local environment[77]. The partial RDF equation may be written as

$$g_{AB}(r) = \frac{\rho_{AB}}{\rho_B} = \frac{dn_{AB}(r)/4\pi r^2 dr}{N_B/V} = \frac{V}{4\pi N_B} \times \frac{dn_{AB}(r)}{r^2 dr} \qquad \text{Eq. 3}$$

where $dn_{AB}(r)$ denotes the number of atoms of species $B$ in a spherical shell of thickness $dr$ at a radial distance $r$ from atoms of species $A$, and $N_B$ is the total number of species $B$ in the simulation box with the volume of $V$. Thus, $g_{AB}(r)$ represents a measure of the local density of species $B$ around species $A$, with respect to the overall density of the former in the whole system.

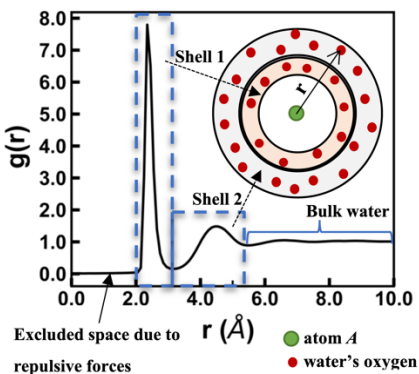

**Figure 3.** Identifying two hydration shells of atom $A$ (green particle) from the minima in the partial RDF of water's oxygen (red particles) around atom $A$.

The hydration shell around a solute atom is defined as a spherical shell centered at the solute's center of mass within which water molecules are packed with a density different than that of bulk water. To identify hydration shells of water around ions and atoms of functional groups on DNDs' surfaces, we use the partial RDF plots of water's oxygen atom with respect to those species (see Figure 3). The region between two successive minima in these plots corresponds to a hydration shell. Different species can have multiple hydration shells, the quantity of which depends on the strength of electrostatic interactions between water and the species.

Another useful information that we can extract from the partial RDF data is the coordination number $n_{AB}(r_1, r_2)$, which is defined as the average number of atoms of species $B$ separated from species $A$ at radial distances between $r_1$ and $r_2$. We calculate $n_{AB}(r_1, r_2)$ by integrating $g_{AB}(r)$ as follows:

$$n_{AB}(r_1, r_2) = \frac{4\pi N_B}{V} \int_{r_1}^{r_2} g_{AB}(r)\, r^2 dr \qquad \text{Eq. 4}$$

The coordination number in the first hydration shell of an atom corresponds to the average number of nearest neighbor water molecules around it.

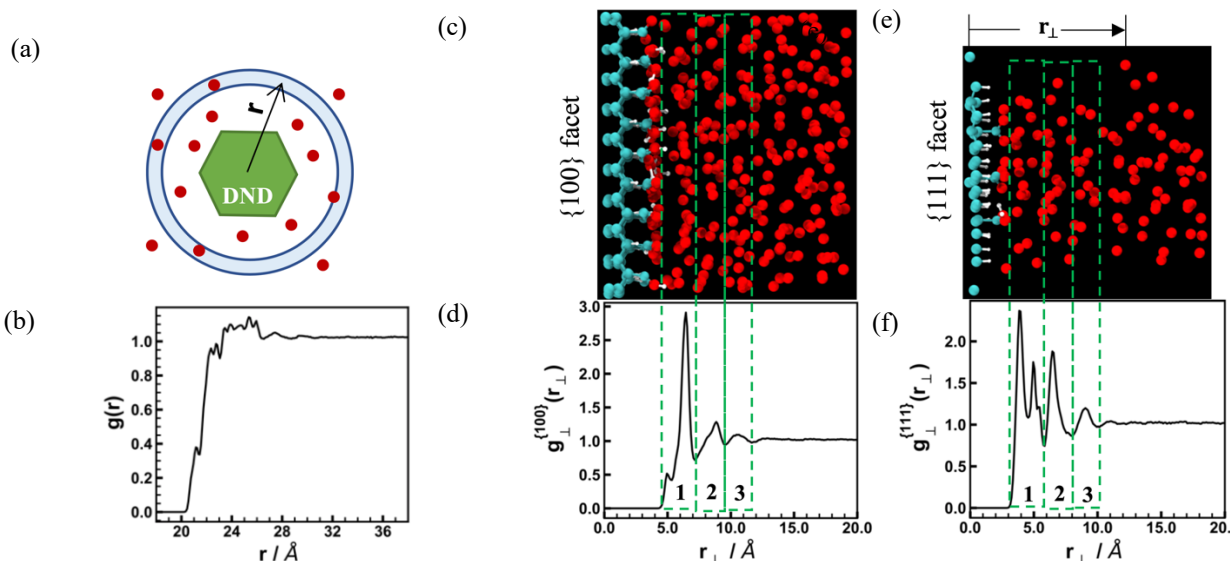

**Figure 4.** (a) Spherical shell around DND to calculate the partial RDF for part (b), (b) the partial RDF of water's oxygen with respect to the centroid of DND–COOH with –84 charges solvated in MgCl$_2$ solution, (c) water's oxygen nearby one of {100} facets of the DND described in part (b), (d) PND of water's oxygen described in part (c), (e) and (f) are the same as (c) and (d) but for one of {111} facets. Particles with cyan, red, and white in (c) and (e) represent carbon, oxygen, and hydrogen atoms, respectively. We have identified three hydration layers at {100} and {111} facets with green dashed rectangles.

### 2.2.2. Perpendicular Number Density

We have found that the pair RDFs can hardly reveal distinguishable hydration shells of DNDs (see Figure 4(b)), in spite of being effective to identify hydration shells of ions. We attribute this issue to highly faceted character of cuboctahedral DNDs in this study. Thus, we instead use normalized Perpendicular Number Density (PND) plots of water to identify hydration layers of DNDs. We define the normalized PND as

$$g^{\xi}_{\perp,\delta}(r_\perp) = \frac{n^{\xi}_{\delta}(r_\perp)/A^{\xi}\, dr_\perp}{N^{\xi}_{\delta}/V^{\xi}} \qquad \text{Eq. 5}$$

where $\delta$ denotes O or H in water, $\xi$ represents either of {100} or {111} facets with surface area $A^{\xi}$, $r_\perp$ is the perpendicular distance of the bin's center from the corresponding facet, $dr_\perp$ is the thickness of bins, $n^{\xi}_{\delta}(r_\perp)$ is the number of $\delta$ atom in a bin at $r_\perp$ from the facet, $V^{\xi}$ represents the total volume of bins and $N^{\xi}_{\delta}$ is the total number of $\delta$ atoms in all bins. We note that the corresponding 3D bins for {100} and {111} facets have cubic and triangular prism shapes, respectively.

In practice, we first calculate $g^{\xi}_{\perp,\delta}(r_\perp)$ separately for each individual facet. Then, we take the average of results for 6 facets of {100} and 8 facets of {111} to obtain the aggregate values of $g^{\xi}_{\perp,\delta}(r_\perp)$ for {100} and {111} facets, respectively. Similar to partial RDFs, $g^{\xi}_{\perp,\delta}(r_\perp)$ also exhibits some fluctuations at short distances and then it tends to unity. The region between two minima in $g^{\xi}_{\perp,0}(r_\perp)$ corresponds to a hydration layer of $\xi$ facet. In Figure 4((d), (f)), we have shown $g^{\{100\}}_{\perp,0}(r_\perp)$ and $g^{\{111\}}_{\perp,0}(r_\perp)$, wherein hydration layers have been specified with green dashed rectangles. Hereafter, if we refer to a property in a particular hydration layer of a DND (e.g., the first hydration layer) without referring to {100} or {111} facets, it means that values of that property have been averaged over the corresponding hydration layer of all 14 facets of the DND.

### 2.2.3. Angular Distribution

The Angular Distribution (AD) is a useful tool to study the preferential orientations of water molecules in a specific hydration layer of DNDs and ions. We are interested in ADs of three different angles $\alpha$, $\beta$, and $\theta$ that are formed between a reference orientation $\vec{R}_{ref}$ and three different vectors attached to a water molecule. The vectors and their associated angles are defined in Table 1. For ADs in the hydration layers of a DND, $\vec{R}_{ref}$ is the outward normal vector to its {100} or {111} facets. However, as far as the preferential orientations of water in the hydration shells of ions are concerned, $\vec{R}_{ref}$ is the vector connecting an ion to the oxygen of the surrounding water.

**Table 1.** Definitions of three different angles related to three different orientations of a water molecule.

| | |
|---|---|
| $\boldsymbol{\alpha}$: $(\widehat{\vec{R}_{ref}, \overrightarrow{OH}})$, where $\overrightarrow{OH}$ is water's OH bond | 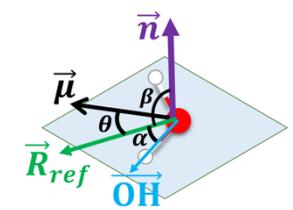 |
| $\boldsymbol{\beta}$: $(\widehat{\vec{R}_{ref}, \vec{n}})$, where $\vec{n}$ is normal to the plane containing water's OH bonds | |
| $\boldsymbol{\theta}$: $(\widehat{\vec{R}_{ref}, \vec{\mu}})$, where $\vec{\mu}$ is water's dipole moment | |

We employ the Distance-dependent Angular Distribution (DAD) to gain insight at a more granular level into the orientational patterns of water around DNDs and ions than what AD can reveal to us. In fact, we use the latter, which is a 1D probability distribution, to obtain the probability of finding water in the first hydration layer of DNDs or ions with a particular orientation. However, the former is a 2D joint probability distribution that yields the probability of finding a water molecule with a particular orientation whose oxygen atom is located at a specific distance away from a reference point. Here, the definition of the reference point and distance measures as well as hydration layers are identical to what presented in Sections 2.2.1 and 2.2.2. Thus, DAD enables us to explore the orientational behavior of water well beyond the first hydration layer of the solute of interest. Computationally, we obtain the AD and DAD from normalized 1D and 2D histograms, respectively.

### 2.2.4. Survival Correlation Function

The Survival Correlation Function (SCF), which was first introduced by Impey *et al.*[78] enables us to identify environments around solutes with distinct dynamic features[79]. The time decay of the SCF measures the probability of finding a solvent molecule in a particular region after a specific elapsed time from its arrival in that region. Thereby, it paves the way to quantify the mean residence time of solvent atoms in hydration layers of solutes. We define the SCF for water molecules in the hydration layer $n$ around a solute as[80]

$$S_n(t) = \langle p_n(t_0)p_n(t_0 + t)\rangle / \langle p_n(t_0)^2\rangle \qquad \text{Eq. 6}$$

where $p_n$ is a binary function that takes the value of 1 if a particular water's oxygen atom resides in the hydration layer $n$ from time $t_0$ to $t_0 + t$, without leaving the region in the interim, otherwise it becomes 0. In addition, $p_n(t_0)^2$ in the denominator represents the number of water molecules existing in region $n$ at time $t_0$, where $t_0$ represents different starting points in time. $\langle \cdots \rangle$ denotes the ensemble average, which means taking the average over all particles of interest and also over all starting times $t_0$. To calculate SCFs, we sample positions of water's oxygen atoms every 0.1 ps during the last 200 ps of the production phase of the MD simulation. The choice of this sampling frequency is based on a trade-off between the computational efficiency and resources on the one hand and capturing the sub-picosecond exchange dynamics of water between hydration shells on the other hand.

The mean residence time $\tau_{res}^{(n)}$ of water in the $n$th hydration layer of DNDs and ions can be obtained by the integration of the functional form of the SCF as[81]

$$\tau_{res}^{(n)} = \int_0^\infty S_n(t)dt \qquad \text{Eq. 7}$$

Our SCF results, which are presented in Section 3.3, show that they do not completely decay to zero by the end of the analysis time, especially for SCFs of the first and second hydration layers. Thus, we have decided to fit the following tri-exponential function to $S_n(t)$ in order to carry out the numerical integration:

$$S_n(t) = \left[A^2 + (1-A^2)\exp\left(-\frac{t}{\tau_1}\right)\right] \times \left[B^2 + (1-B^2)\exp\left(-\frac{t}{\tau_2}\right)\right] \times \exp\left(-\frac{t}{\tau_3}\right) \qquad \text{Eq. 8}$$

We have chosen this functional form for the SCF for three reasons. First, other studies have shown that the decay behavior of the SCF for water in constrained environments cannot be described by a simple exponential function.[82] Secondly, this functional form can somehow reflect the multi-scale relaxation behavior of the dynamics of water in hydration shells of solute, which has been reported in the literature[83,84]. Third, we have found that this functional form fits very well to the SCF of water in the first and the second hydration layers of DNDs. For the third hydration layer, since SCF decays to zero within the time window of our analysis, we can readily carry out the integral using some numerical recipe.

## 3. Results and Discussions

### 3.1. Hydration layers

In Figure 5, we have compared the effects of KCl and MgCl$_2$ solutions on $g_\perp(r_\perp)$ of water and also partial RDFs of water around polar surface groups of DND–COOHs. In these plots, $g_\perp^{\{100\}}(r_\perp)$ and $g_\perp^{\{111\}}(r_\perp)$ correspond to, respectively, averaged values over six {100} and eight {111} facets. $g_\perp(r_\perp)$ of water hydrogen in Figure 5(a-d) exhibit three distinct changes, as DND–COOH acquires more negative charges. Firstly, the second peak's intensity reduces, while the first peak grows more in intensity and gets closer to facets of the more negatively charged DND–COOH. However, the increase in the first peak's intensity is more pronounced in MgCl$_2$ solution than in KCl solution. Thus, it indicates that the local density of water hydrogen in the first hydration layer of the charged DND–COOH is higher in the former than that in the latter. Secondly, irrespective of the type of the solution and the number of negative charges on DND–COOH, the aforementioned first peak is more intense at {111} facets than that at {100} facets.

As we mentioned above, water hydrogens are found closer to DND–COOH surfaces with more negative charges. It originates from –COO$^-$ surface groups that are HB acceptors from the interfacial water. In Figure 5(e and g), the partial RDFs of water atoms around O atom of –COO$^-$ groups are reported, which corresponds to KCl and MgCl$_2$ solutions, respectively. We observe narrow sharp peaks close to the surface in water hydrogen's partial RDFs, although MgCl$_2$ solution is associated with more pronounced first peak intensity than KCl solution.

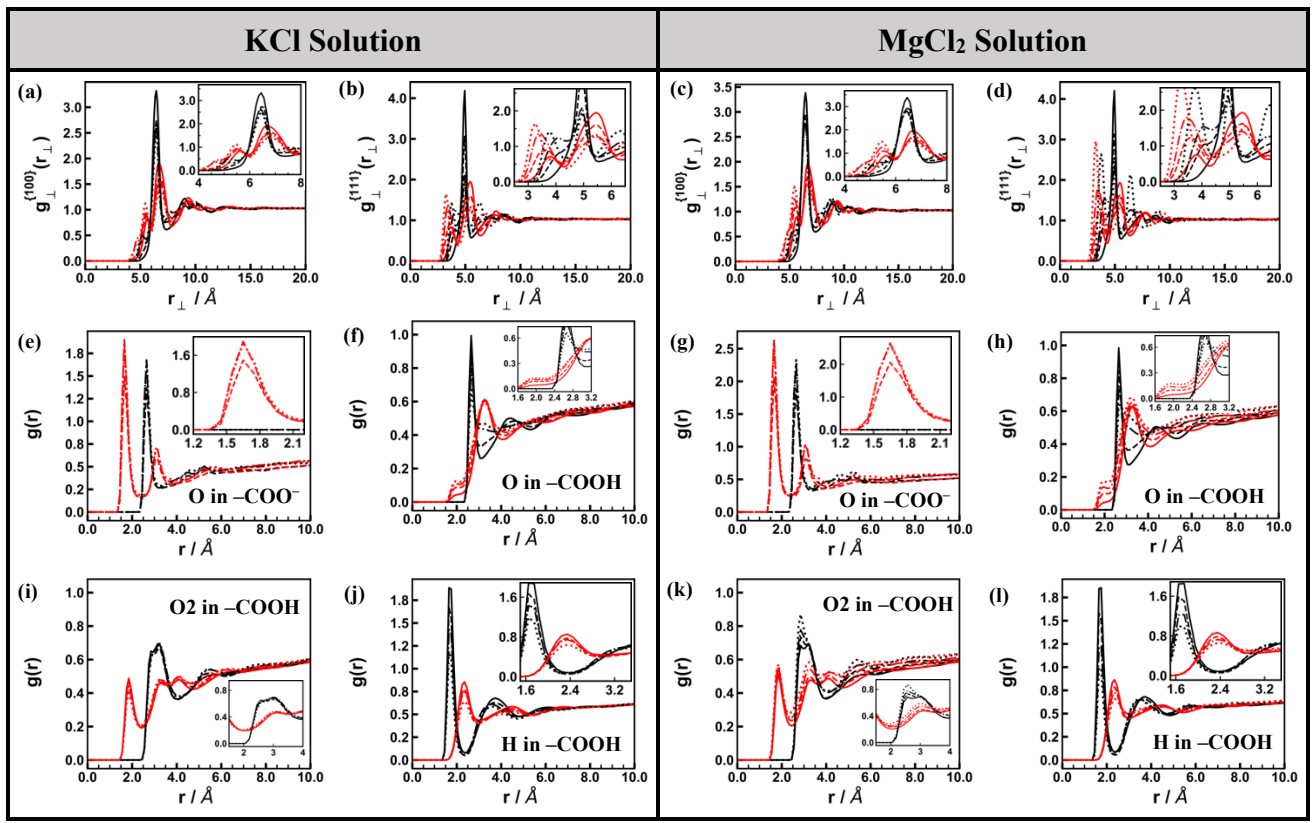

**Figure 5.** Comparing KCl and MgCl$_2$ solutions of DND–COOH: (a)-(d) PND plots of water oxygen and hydrogen atoms at {100} and {111} facets, (e)-(l) partial RDFs of water oxygen and hydrogen atoms around constituent atoms of –COOH and –COO$^-$ polar groups on surfaces of the DND. O and O2 in –COOH group are, respectively, oxygen atom of the hydroxyl (OH) group and oxygen atom doubly-bonded to carbon atom. Black and red colors correspond to oxygen and hydrogen atoms of water, respectively. Furthermore, solid, dashed, dash-dotted, and dotted lines represent DNDs with 0, –28, –56, and –84 charges, respectively. Negative charges come from –COO$^-$ groups.

Partial RDFs of water atoms around O and H atoms of –COOH group are presented in Figure 5(f and j) for KCl and in Figure 5(h and l) for MgCl$_2$ solution. In these plots, we can see a shoulder at $r \approx 2$ Å in the RDF of water hydrogen around O atom (i.e., the oxygen atom of the hydroxyl (OH) group) and a pronounced peak at $r \approx 1.7$ Å in the RDF of water oxygen around H atom (i.e., the hydrogen atom of the hydroxyl (OH) group). Thus, it leads us to conclude that the hydroxyl portion of –COOH primarily acts as a hydrogen donor in HBs with the nearby water. Nevertheless, the aforementioned shoulder suggests that some few water molecules still attempt to donate a hydrogen atom to form an HB with the OH atom of –COOH group. In addition, this shoulder appears to be more pronounced in MgCl$_2$ solution than KCl solution at higher surface charges. In Figure 5(i and k), the partial RDFs of water around O2 atom of –COOH group (oxygen atom doubly-bonded to carbon atom) are compared between KCl and $MgCl_2$ solutions, respectively. We can see in these RDFs, similar to the aforementioned RDFs, a higher intensity of peaks for both water hydrogen and oxygen in $MgCl_2$ solution compared to KCl solution.

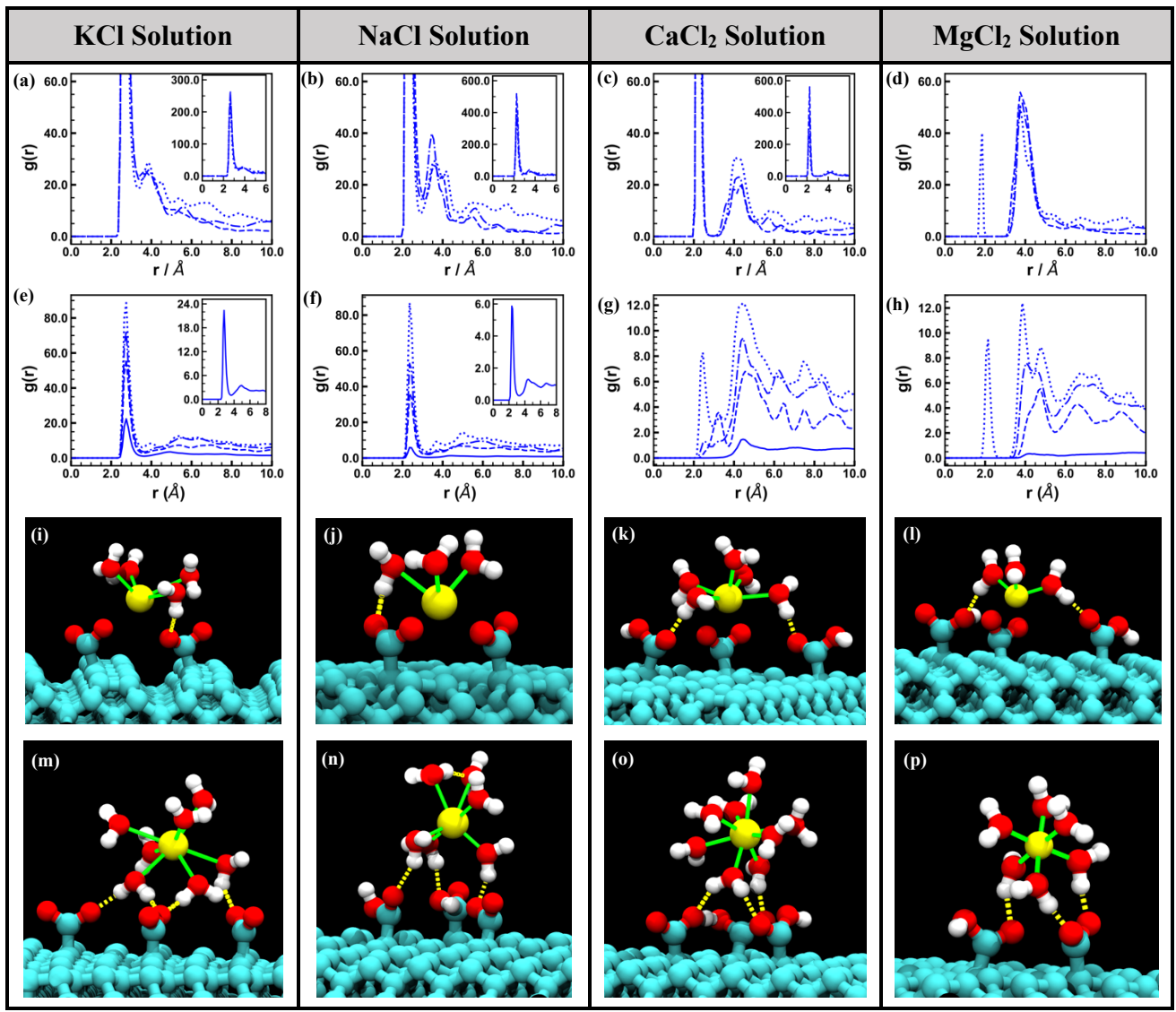

**Figure 6.** Comparing four solutions of DND–COOH: (a)-(d) partial RDFs of the constituent cation of solvated salts around O in –COO⁻, (e)-(h) same as previous ones but for O2 in –COOH, (i-l) snapshots from MD trajectories to show Contact Ion-Pair (CIP) association of the constituent cation of solvated salts with –COO⁻ anion on DND–COOH with –84 charges, (m-p) same as (i-l) but for Solvent-shared Ion-Pair (SIP) association. Line styles in (a)-(h) are the same as those in Figure 5. In snapshots, particles with cyan, red, white, and yellow colors represent carbon, oxygen, hydrogen, and the cation, respectively; dotted yellow lines show HBs between water and polar groups; and green lines are included just for visual aid.

We observed above that $MgCl_2$ solutions are associated with a richer water organization in hydration layers of negatively charged DND–COOHs relative to KCl solutions. It could be due to different interactions between the specific cation solvated in the solution and –COO⁻ anions on DND–COOH surfaces. Furthermore, the negative partial charges on O2 that is doubly bonded to C in –COOH group can also induce some electrostatic interactions with cations. We thus aim at investigating aforementioned interactions by analyzing the partial RDF plots of various cations around O in –COO⁻ group as well as O2 in –COOH, which are shown in Figure 6. Peak positions in ranges of 1.85–2.65 Å and 3.45–4.35 Å for different cations correspond to, respectively, their Contact Ion-Pairing (CIP) and Solvent-shared Ion-Pairing (SIP) with –COO⁻ and also –COOH groups. The former arises from two ions coming to a close contact, whereas the latter is formed when two ions are separated by one solvent (i.e., water) molecule.[85] There is also a third peak, although very shallow, in some of the partial RDF plots that points to the formation of Solvent-Separated Ion-Pairs (2SIP) between some cations and –COO⁻ anion and also –COOH group. In 2SIP complexation, two ions are completely separated from each other by their first solvation (hydration) shell.[86–91]

Aforementioned peak positions and their intensities for ion-pairing of cations and –COO⁻ anion are listed in Table 2. In addition, we have illustrated snapshots of sampled atomic configurations from our MD simulations in Figure 6 for adsorbed cations on a surface of DND–COOH with –84 charges. Figure 6(i-l) and Figure 6(m-p) show, respectively, CIP and SIP associations of cations with –COO⁻. Interestingly, cations in the former are surrounded by fewer number of water molecules than those in the latter. It implies that cations become partially dehydrated upon their adsorption in the CIP mode.

**Table 2.** The position and intensity of the first three peaks in the partial RDFs of cations around –COO⁻ surface-bound anion on DND–COOH. The first, second, and third peaks correspond to, respectively, CIP, SIP, and 2SIP associations of cations with –COO⁻ anion.

| Ion pair | DND's surface charge | CIP | | SIP | | 2SIP | |
|---|---|---|---|---|---|---|---|
| | | $r$ | $g(r)$ | $r$ | $g(r)$ | $r$ | $g(r)$ |
| **K⁺ – COO⁻** | q = –28 | 2.65 | 257.9 | 3.65 | 25.3 | – | – |
| | q = –56 | 2.65 | 265.2 | 3.85 | 25.7 | 5.55 | 11.3 |
| | q = –84 | 2.65 | 238.6 | 3.85 | 28.7 | 5.35 | 14.7 |
| **Na⁺ – COO⁻** | q = –28 | 2.25 | 518.2 | 3.55 | 27.4 | 5.45 | 7.7 |
| | q = –56 | 2.25 | 425.4 | 3.45 | 39.9 | 5.65 | 10.9 |
| | q = –84 | 2.25 | 423.6 | 3.55 | 29.2 | 5.45 | 13.2 |
| **Ca²⁺ – COO⁻** | q = –28 | 2.25 | 561.2 | 4.35 | 19.7 | – | – |
| | q = –56 | 2.25 | 378.3 | 4.25 | 23.0 | 5.75 | 7.3 |
| | q = –84 | 2.25 | 256.4 | 4.05 | 30.7 | 5.75 | 9.9 |
| **Mg²⁺ – COO⁻** | q = –28 | – | – | 3.85 | 55.8 | – | – |
| | q = –56 | – | – | 3.85 | 51.6 | 5.25 | 7.0 |
| | q = –84 | 1.85 | 40.1 | 3.75 | 50.2 | 5.25 | 8.5 |

We note in Table 2 that the corresponding peak intensities to CIP complexation of Na⁺ with –COO⁻ at all charge values of DND–COOH are almost twice as large as those of K⁺ cation. It implies that the former has a higher binding affinity to –COO⁻ than the latter. This observation corroborates prior studies on the ion-pairing of these two cations with the carboxylate anion that exists in different physiological environments (e.g., as a free intracellular anion or as an adduct on protein surfaces).[68,86,92] Collins explained the aforementioned difference via the so-called "law of matching water affinity"[83]. This empirical rule states that oppositely charged ions with matching water affinities form CIPs in the solution, whereas the ones with differing water affinities tend to stay apart. In other words, like attracts like.

The water affinity of an ion is a function of its surface charge density, which in turn determines whether it is strongly hydrated (kosmotrope) or weakly hydrated (chaotrope). Small ions with high charge densities are strongly hydrated and thus have high water affinities. In contrast, ions with low water affinities are large monovalent ions that are weakly hydrated. Thus, according to the law of matching water affinity, CIP associations between kosmotrope–kosmotrope and chaotrope–chaotrope are much more likely to form than that between kosmotrope–chaotrope. Collin argues that the latter is less likely to be formed due to energetic considerations. That is, the energy gain resulting from the kosmotrope–chaotrope CIP complexation does not compensate the energy cost of partially dehydrating the strongly hydrated kosmotrope.

To characterize the water affinity of ions, various techniques have been suggested, for example, measuring the Apparent Dynamic Hydration Number (ADHN) or the Jones–Dole viscosity B coefficient of ions in salt solutions[92]. The former is a measure of the number of tightly bound water molecules to an ion as it diffuses through a medium and thus is different from its static coordination number obtained from RDF plots. The latter yields a metric for the strength of water–ion interactions with respect to those of water–water in bulk solution. In Table 3, values of ADHN and B coefficient for the relevant ions in our study are presented. The more positive is the B coefficient of an ion, the more strongly hydrated it is and a higher water affinity it has. In addition, an ion with a higher value of ADHN binds water more tightly to its immediate vicinity and hence has a higher water affinity. Thus, based on the values in Table 3, $Mg^{2+}$ is the strongest hydrated ion in our study, while $K^+$ and $Cl^-$ ions are the weakest hydrated ions.

Table 3. Characteristics of ions determining their hydration strength.[78]

| | Ion | Jones-Dole viscosity B Coefficient | ADHN | Ionic radius (Å) | Surface charge density (C/m²) |
|---|---|---|---|---|---|
| Strongly Hydrated | $HCOO^-$ | 0.052 | 2.0 | | |
| | $Na^+$ | 0.086 | 0.22 | 1.02 | 1.06 |
| | $Ca^{2+}$ | 0.285 | 2.1 | 1.00 | 2.24 |
| | $Mg^{2+}$ | 0.385 | 5.9 | 0.72 | 4.3 |
| Weakly hydrated | $Cl^-$ | –0.007 | 0.0 | 1.81 | 0.39 |
| | $K^+$ | –0.007 | 0.0 | 1.38 | 0.59 |

We proceed to interpret the adsorption of cations onto surfaces of charged DND–COOH, using the law of matching water affinity and characteristics of ions presented in Table 3. First of all, B coefficient of $Na^+$ is the closest to that of $-COO^-$, compared with other cations. Thus, the former has the highest tendency to form CIP association with the latter[93]. Secondly, the formation of $Ca^{2+}$–$COO^-$ CIP complex can be justified by the closeness of their ADHN values[94]. That is, they both tightly bind about two water molecules to their immediate vicinity. Thirdly, the significant mismatch between water affinities of $Mg^{2+}$ and $-COO^-$ can explain why $Mg^{2+}$ only forms SIP rather than CIP association with $-COO^-$ at moderate concentrations of these ions. This condition corresponds to –56 or lower number of negative charges, which arises from $-COO^-$, on DND–COOH. Previous studies have also cast a strong doubt on any formation of CIP association between $Mg^{2+}$ cation and –$COO^-$ anion, in particular at low to moderate concentrations.[95]

The law of water affinity, however, fails to explain two patterns we have observed. First, the corresponding peak intensities to CIP associations between $Ca^{2+}$ and –$COO^-$ significantly drop with increasing concentrations of $Ca^{2+}$ and –$COO^-$, while those corresponding to SIP associations increase (see Table 2). Second, the CIP complexation of $Mg^{2+}$–$COO^-$ emerges, when DND–COOH acquires –84 net charges (that is, at higher concentrations of these ions). These observations are less likely due to any artifacts introduced by forcefield parametrizations of ions or the water model used in our MD simulations, because of two reasons. First, our observations corroborate the theoretical investigation performed by Fedotova and Kruchinin on the adsorption of the same cations as ours onto –$COO^-$ anion[96,97]. They reported that the stability of ion-pairs between the cations and –$COO^-$ decreases with increasing ion concentrations. Second, CIP complexations of $Mg^{2+}$–$COO^-$ at high concentrations have also been observed in previous experimental[93] and MD simulations with both polarizable and non-polarizable water models[41,98]. In particular, Wang *et al.* have reported[99] the formation of bidentate structures for CIP complexes of $Mg^{2+}$–$COO^-$ in supersaturated droplets of magnesium acetate ($CH_3COO^-$), similar to what we have shown in Figure 6(l). In addition, Semmler *et al.* have provided the stability constants for the formation of magnesium acetate complexes as a function of temperature and concentrations of $Mg^{2+}$ and $CH_3COO^-$ ions[100].

We hypothesize that $Mg^{2+}$ cations form CIP associations with –$COO^-$ at high concentrations due to entropic gains. Our hypothesis is based on two hydration characteristics of $Mg^{2+}$ cation. First, it is a strongly hydrated cation, as evidenced by its high B coefficient and ADHN values in Table 3. Interestingly, its ADHN of 5.9 is almost identical to its static hydration number of 6, meaning water molecules are tightly bound to the first hydration shell of $Mg^{2+}$. Thus, in its fully hydrated state, i.e., $[Mg(H_2O)_6]$, it appears as a relatively large sphere with $Mg^{2+}$ at the center and 6 coordinating water molecules decorated at the outer surface. Second, $Mg^{2+}$, as we will see in the ensuing sections, imparts a highly ordered structure in its surrounding water. It implies that higher adsorption of $Mg^{2+}$ cations leads to a higher ordered structure in the hydration layers of charged DND–COOH. On the one hand, the free space, which is created as a result of partially dehydration of $Mg^{2+}$ cations upon CIP association, facilitates the adsorption of more $Mg^{2+}$ cations to hydration layers of DND–COOH. On the other hand, the resulting entropic gain obtained from the higher order induced by $Mg^{2+}$ compensates the energy that was spent to partially dehydrate the strongly hydrated $Mg^{2+}$ in the process of CIP complexation.

We end the discussion of cation adsorptions on DND–COOH by explaining why $K^+$ and to a lesser degree $Na^+$, but not $Mg^{2+}$ or $Ca^{2+}$, form CIP associations with O2 in –COOH, particularly when the DND is neutral (see Figure 6(e-h)). We have not been able to find the B coefficient or ADHN values for O2 in the literature to compare its water affinity with that of cations. However, we can obtain a qualitative measure of the water affinity of O2. The first peak intensity in partial RDF of water hydrogen around O2 in –COOH is 4-5 times smaller than that around O in –$COO^-$ (see Figure 5(e, i) and also Figure 5(g, k)). Thus, it implies that the water affinity of O2 is by far less than that of O in –$COO^-$ and in turn is less than that of $Mg^{2+}$ and $Ca^{2+}$ cations. Thus, according to the law of matching water affinity, the weakly hydrated O2 cannot pair with these divalent cations. In fact, strongly hydrated $Mg^{2+}$ and to a lesser degree $Ca^{2+}$ have significantly low mobilities in water[96]. It arises from their tightly bound hydrating water being hydrogen bonded to the neighboring water molecules. Thus, their diffusion in the solution encounters significant frictional forces[83,101], which cannot be overcome by weak electrostatic potential of O2 atoms. In contrast, $K^+$ has a much higher mobility due to its labile hydration shell, as evidenced by its ADHN value of 0 (see Table 3). Thus, even the weak electrostatic potential of O2 is sufficient to attract certain amounts of $K^+$ to facets of the uncharged DND–COOH. The mobility of $Na^+$ is a bit smaller than that of $K^+$. Thus, it is adsorbed less significantly than $K^+$, as evident in insets to Figure 6(e-g)).

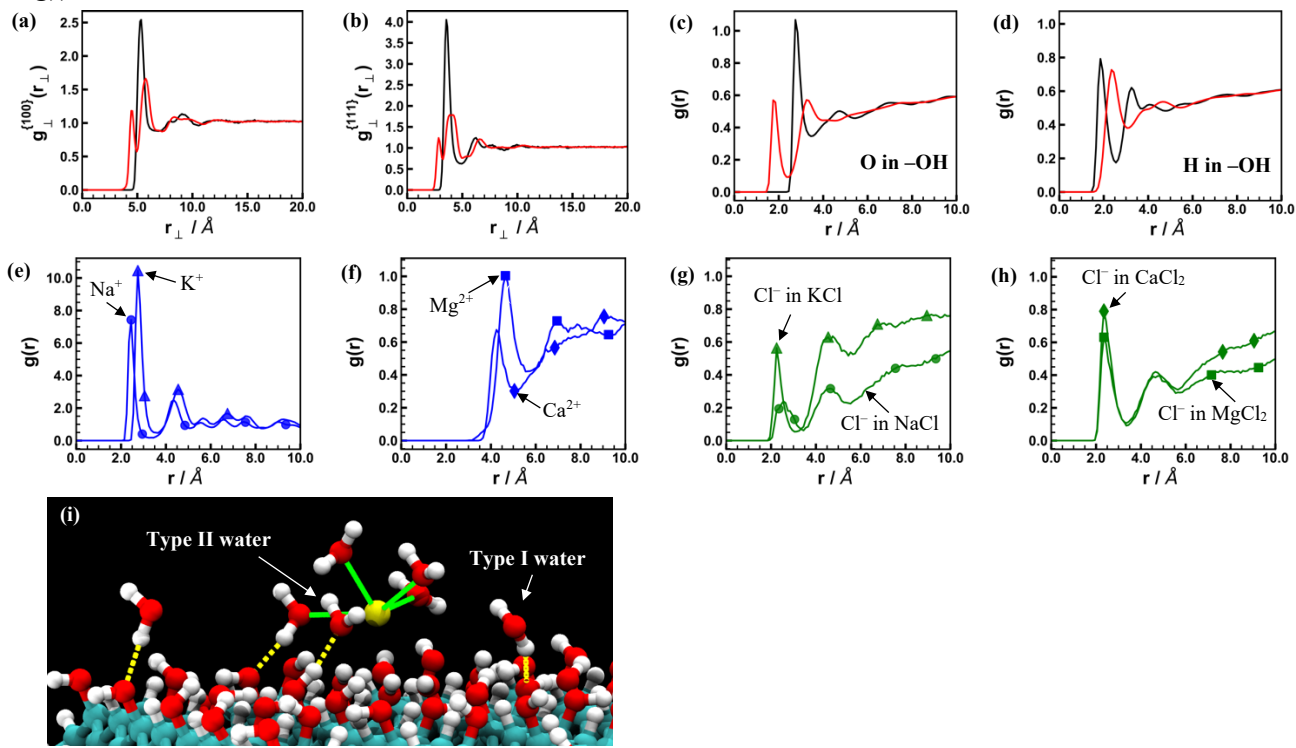

**Figure 7.** (a)-(b) PND plots of water oxygen and hydrogen atoms at {100} and {111} facets for DND–OH in KCl solution, (c)-(d) partial RDFs of water oxygen and hydrogen atoms around constituent atoms of –OH on surfaces of DND–OH in KCl solution, (e)-(f) partial RDFs of cations around O atom of –OH group in four different salt solutions of DND–OH, (g)-(h) same as (e)-(f) but for $Cl^−$ around H atom of –OH group, (h) a snapshot from MD trajectories of DND–OH in KCl solution showing a portion of the first hydration layer of the DND at {100} facet. Black, red, blue, and green lines in plots correspond to water oxygen, water hydrogen, cation, and anion, respectively. Color schemes in the snapshot are the same as those in Figure 6. The snapshot shows $K^+$–OH CIP association as well as two types of water: type I and type II. The former is a HB donor, while the latter is a HB acceptor.

We proceed by exploring the hydration structure around DND–OH. Density plots of water perpendicular to its {100} and {111} facets, solvated in KCl salt solution, are shown in Figure 7(a, b). In addition, the partial RDFs of water atoms around constituent atoms of the surface hydroxyl group are reported in Figure 7(c, d). Features of these plots are similar to those of the corresponding plots for other salt solutions. At both types of facets, water oxygen density exhibits a major peak followed by a second shallow peak and subsequently a small bump, although the first peak of {111} facets is slightly sharper.

From the peak positions of partial RDFs in Figure 7(c, d), we infer that the surface hydroxyl group can act as both a HB donor to and a HB acceptor from the interfacial water. More specifically, the first peak position in $g_{O-HW}(r)$ (red line in Figure 7(c)) corresponds to the average interatomic distance between water hydrogen (HW) and O in –OH group as an HB acceptor from water. Furthermore, the first peak position in $g_{H-OW}(r)$ (black line in Figure 7(d)) represents the average separation distance between water oxygen (OW) and H in –OH as an HB donor to water. Thus, the first peak of red and black lines in both $g_\perp^{\{100\}}(r_\perp)$ and $g_\perp^{\{111\}}(r_\perp)$ correspond to water molecules that are HB donors and HB acceptors from –OH groups, respectively. The snapshot in Figure 7(e) illustrates these two groups of water in the first hydration layer of DND–OH in KCl solution. In addition, it shows the CIP association between an adsorbed $K^+$ cation and the –OH group, where the cation is partially dehydrated as a result of the association.

The partial RDFs of monovalent and divalent cations around O of –OH group, which has negative partial charges, are demonstrated in Figure 7(e) and Figure 7(f), respectively. First peaks in the former point out the CIP association as the preferential adsorption of $K^+$ and $Na^+$ onto the hydroxyl group. In contrast, the latter leads us to conclude that $Mg^{2+}$ and $Ca^{2+}$ cations prefer to make SIP complexation with –OH group. The same argument that we presented above to explain the association of these cations with –COOH group on the uncharged DND–COOH can also be applied here.

We report some interesting patterns in the adsorption of $Cl^-$ anions from different chloride salt solutions of cations onto facets of DND–OH. First, consider Figure 7(g) and Figure 7(h) that illustrate the partial RDFs of $Cl^-$ anion around H of –OH group. The former compares the partial RDFs obtained from KCl and NaCl solutions, while the latter reports those of $CaCl_2$ and $MgCl_2$ solutions. The position of the first peak in these plots indicates the formation of the CIP association between H of –OH group and $Cl^-$ anion. The second pattern worth noting is the different adsorption behavior of $Cl^-$ anions on facets of DND–OH. As it can be seen in Figure 8, $g_\perp^{\{111\}}(r_\perp)$ plots for $Cl^-$ exhibit pronounced first peaks and even second peaks in the case of $CaCl_2$ and $MgCl_2$ solutions, both of which are absent in $g_\perp^{\{100\}}(r_\perp)$ plots. It points out that there is a much higher probability for $Cl^-$ anions to be adsorbed on {111} facets than on {100} facets. Furthermore, the relative position of first peaks in $g_\perp^{\{111\}}(r_\perp)$ for $K^+$ and $Cl^-$ indicates that {111} facets have higher affinities for the former than the latter. In contrast, {111} facets have higher affinities for $Cl^-$ than the corresponding cation in NaCl, $CaCl_2$, and $MgCl_2$ solutions.

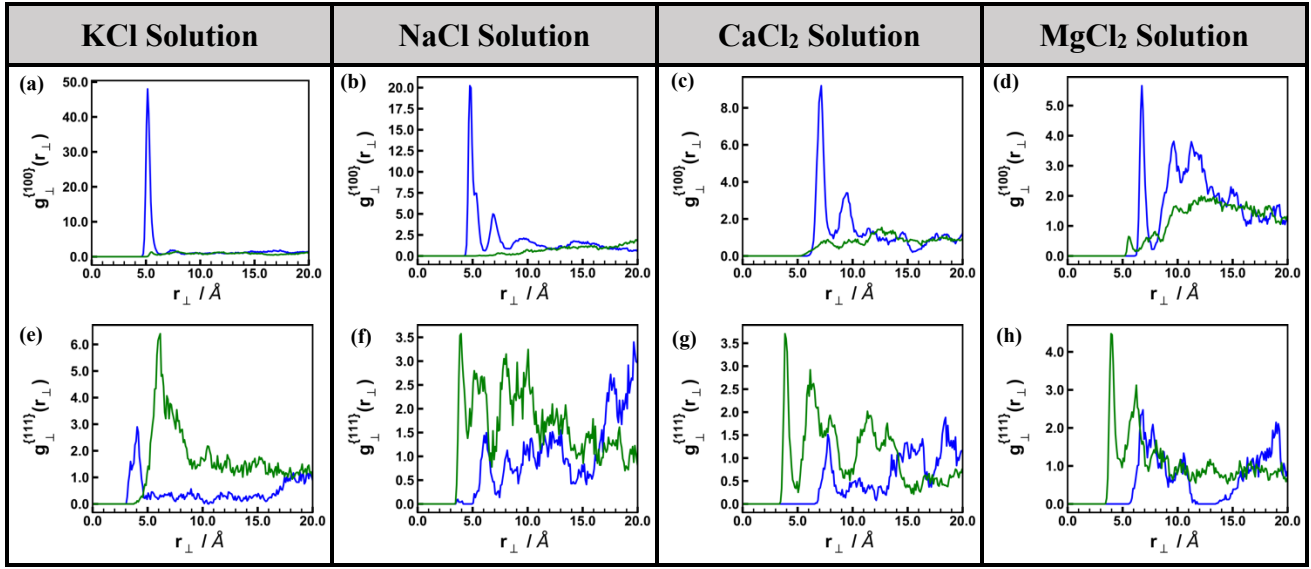

**Figure 8.** Comparing four different salt solutions of DND–OH: (a)-(d) PNDs of the constituent cation and anion of solvated salts with respect to {100} facets, (e)-(h) the same as those of (a)-(d) but with respect to {111} facets. Blue and green lines are for cation and anion, respectively.

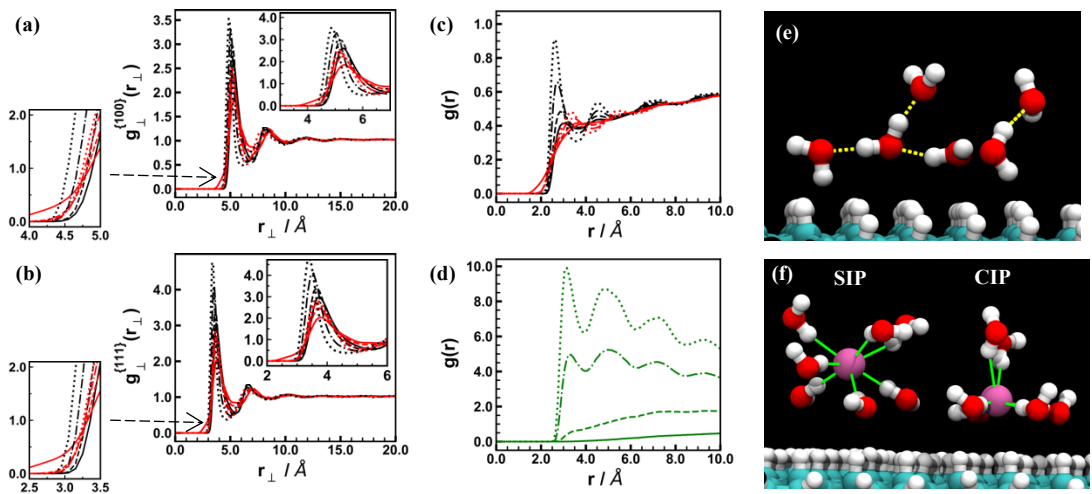

**Figure 9.** (a)-(b) PND plots of water oxygen and hydrogen atoms at {100} and {111} facets for DND–H in NaCl solution, (c) partial RDFs of water oxygen and water hydrogen around H atom on surfaces of DND–H in NaCl solution, (d) the same as (c) but for Cl⁻ anion around H atom, (e) a snapshot from MD trajectories of the neutral DND–H in NaCl solution showing a portion of the first hydration layer of the DND, (f) the same as (e) but for DND–H with +84 charges. Black, red, and green lines in plots correspond to water oxygen, water hydrogen, Cl⁻ anion, respectively. Color schemes in the snapshot are the same as those in Figure 6, except for purple-colored particle that represents Cl⁻ anion. Furthermore, (e) show CIP and SIP associations of Cl⁻ with surface H atoms.

We now turn our attention to the density plots of water and ions in Figure 9 that reveal interesting water layering and solvent adsorption patterns around DND–H. Only results of DND–H solvated in NaCl solution are included, as other salt solutions exhibit similar features. Two trends emerge in $g_\perp(r_\perp)$ plots with increasing positive charges on surfaces of DND–H (see Figure 9(a)-(b)). On the one hand, the first peak for water oxygen gets closer to both {100} and {111} facets. On the other hand, as it can be seen in the insets to the left of Figure 9(a)-(b), the shoulder for water hydrogen near the facets of the uncharged DND–H disappears in the case of charged DNDs. It implies that water hydrogen moves away from surfaces of positively charged DND–H. These observations become clearer by referring to snapshots of selected water molecules nearby facets of DND–H with zero and +84 net charges that are shown respectively in Figure 9(e)-(f). In particular, we can see in Figure 9(e) that two water molecules have one of their hydroxyls (OH) oriented toward the surface of the uncharged DND–H. In contrast, all water molecules in Figure 9(f) have pointed their OH bonds away from the surface of DND–H with +84 net charges.

The charged DND–H particle is decorated with H hydrogen atoms with positive partial (fractional) charges, which altogether give rise to a specific amount of net positive charges for the whole DND. Consequently, these hydrogen atoms attract Cl⁻ anions to surfaces of the positively charged DND–H. The partial RDF plots for H–water oxygen (OW) and H–Cl⁻ are demonstrated in Figure 9(c)-(d). They lead us to infer that the adsorbed Cl⁻ anions form CIP and SIP associations with H atoms on surfaces of DND–H with +56 and +84 net charges. These associations correspond to the first and second peaks in Figure 9(d), respectively. Our rationale for appointing the aforementioned first peak to CIP association is its closeness to the first peak in $g_{H-OW}(r)$ shown in Figure 9(c), at each corresponding DND's surface charge. The snapshot in Figure 9(f) also provides additional evidence for the formation of CIP and SIP associations between H on DND and Cl⁻. The formation of H–Cl⁻ CIP complex can be explained by the law of matching water affinity. Since H and Cl⁻ are both weakly hydrated, thus their water affinities match and they form CIP complexations. In fact, it is energetically more favorable for water molecules to leave the hydration shells of these weakly hydrated species and form stronger water-water association (i.e., HBs) in the bulk.[103,104]

The arrangement of water molecules in Figure 9(e) is in accord with the generally accepted belief that hydrogenated diamond surfaces are hydrophobic. It implies that they do not form HBs with surrounding water molecules. Thus, some interfacial water molecules end up having some dangling hydroxyls pointed toward the DND's surface that do not participate in any hydrogen bonding. However, it appears that the water organization at the interface with positively charged DND–H is completely different from that of the neutral DND–H in two ways. On the one hand, the positively charged surfaces push away the aforementioned dangling hydroxyls and promote them to form HBs with the neighboring water molecules. On the other hand, the adsorbed $Cl^-$ anions in hydration layers of positively charged DND–H influence the arrangements of water molecules, as shown in Figure 9(h). In particular, $Cl^-$ anions can form labile HBs with the interfacial water, which means they quickly change their HBs partners in water[83]. Therefore, if they get adsorbed at high concentrations, which corresponds to high positive charges on DND–H, they can subsequently disrupt the interfacial water's HB network.

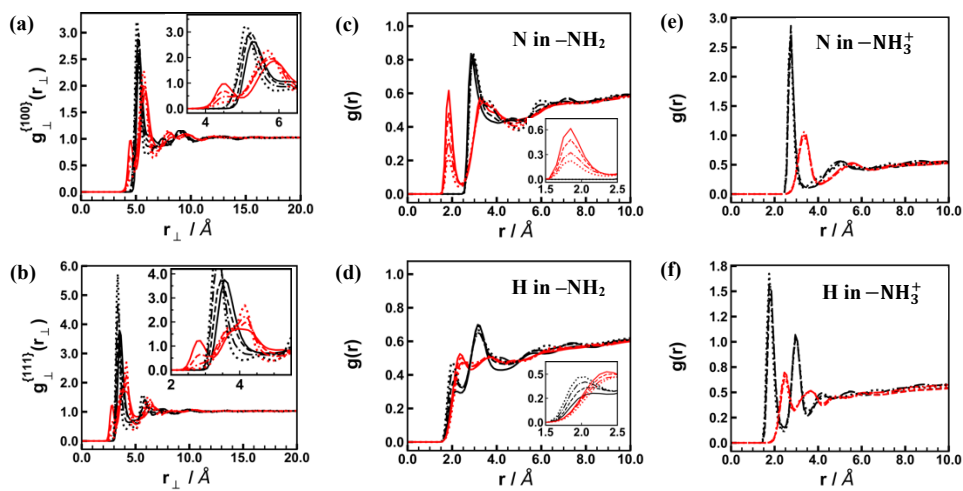

**Figure 10.** (a)-(b) PND plots of water oxygen and hydrogen atoms at {100} and {111} facets for DND–$NH_2$ in NaCl solution, (c)-(f) partial RDFs of water oxygen and water hydrogen around constituent atoms of –$NH_2$ and –$NH_3^+$ groups on DND–$NH_2$ in NaCl solution. Black and red colors correspond to oxygen and hydrogen atoms of water, respectively. Furthermore, solid, dashed, dash-dotted, and dotted lines represent DNDs with 0, +28, +56, and +84 charges, respectively. Positive charges come from –$NH_3^+$ group.

Next, we analyze PND plots for water around DND–$NH_2$ solvated in NaCl solution that are shown in Figure 10(a)-(b). They demonstrate more complex structured water compared with those of DND–H, owing to the existence of –$NH_2$ and –$NH_3^+$ polar surface groups; the latter only exist on the charged DND–$NH_2$. We observe a relatively shallow first peak for water hydrogen atoms close to both {100} and {111} facets of the neutral DND–$NH_2$, whose surfaces are covered by –H and –$NH_2$ groups. We attribute the first peak to water hydrogens that are donated in HBs between water and N atom of –$NH_2$ groups, as it is also evident in Figure 11(d). However, this peak disappears in the case of DND–$NH_2$ with +84 charges, whereas the first peak of water oxygen density therein is closer to the surface than that of the neutral DND. It points to the increasing propensity of the interfacial water of the more positively charged DND–$NH_2$ to orient its OH bonds away from DND's surfaces so as to accept an HB from –$NH_3^+$ groups. The snapshot in Figure 11(f), obtained from MD trajectories, more vividly illustrates HBs formed between water and $-NH_3^+$ groups on a facet of DND–$NH_2$ with +84 charges.

The partial RDFs of water around constituent atoms of –$NH_2$ and $-NH_3^+$ groups are shown in Figure 10(c)-(f), which provide additional support for aforementioned observations. First, the inset to Figure 10(c) shows that the intensity of the first peak in the partial RDF of water hydrogen around N in –$NH_2$ group diminishes, as DND–$NH_2$ becomes more positively charged. Second, we observe in Figure 10(d) that a peak emerges in the partial RDF of O in water around H in –$NH_2$, when the surface charges increase. Third, we note a pronounced peak in the partial RDF of O in water around H in $-NH_3^+$ group, which is shown in Figure 10(f).

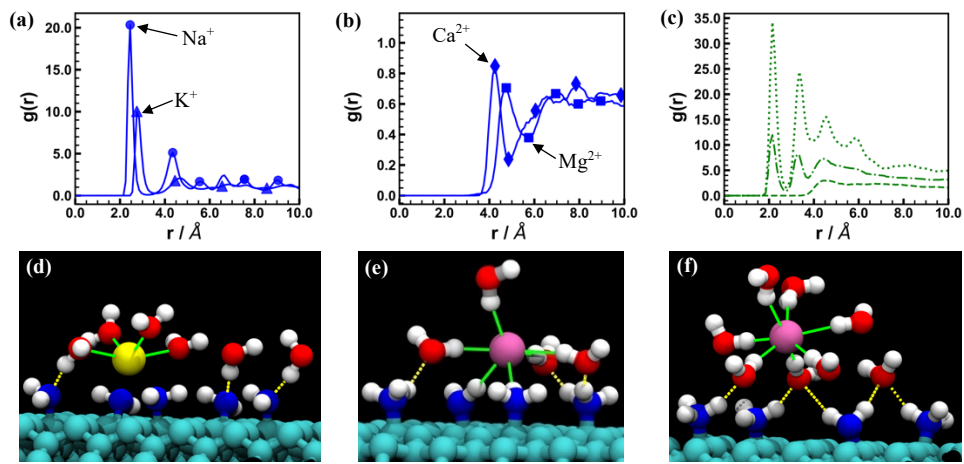

**Figure 11.** (a)-(b) partial RDFs of cations around N atom of –$NH_2$ group of the neutral DND–$NH_2$ in four different salt solutions, (c) partial RDFs of $Cl^-$ anion around H atom of $-NH_3^+$ group on positively charged DND–$NH_2$ in NaCl solution, (d) a portion of the first hydration layer of the neutral DND–$NH_2$ in NaCl solution, (e)-(f) the same as (d) but for DND–$NH_2$ with +84 charges. Line styles in plots are the same as those in Figure 10. Particles with cyan, blue, red, white, yellow, and purple colors in snapshots represent carbon, nitrogen, oxygen, hydrogen, $Na^+$, and $Cl^-$, respectively. Furthermore, (d) and (e) show, respectively, CIP associations of $Na^+$–$NH_2$ and $Cl^-$–$NH_3^+$, while (f) illustrates SIP association of $Cl^-$–$NH_3^+$. For more clear visualization, only a handful of amine groups are included, and hydrogen atoms bonded to surface carbon are not shown.

The adsorption of ions on surfaces of DND–$NH_2$ also demonstrates interesting patterns. Figure 11(a) and Figure 11(b) show partial RDFs of, respectively, monovalent and divalent cations around N in –$NH_2$ groups on surfaces of the neutral DND–$NH_2$. Since the N atom has some negative partial charges, it attracts some cations onto facets of the DND. However, the amount of adsorbed divalent cations is negligible compared with that of monovalent cations. Moreover, it appears in the RDF plots that $Na^+$ forms both CIP (see snapshot in Figure 11(d)) and SIP associations with –$NH_2$ groups, while $K^+$ only forms the former. These two observations can be explained in terms of the water affinity of cations and also the strength of N–cation electrostatic interactions. The detailed mechanism is similar to what we explained before for O2–cation interactions in the uncharged DND–COOH solution.

Partial RDFs in Figure 11(c) indicate more pronounced adsorption of $Cl^-$ anion from NaCl solution on surfaces of charged DND–$NH_2$ covered with 56 or 84 $-NH_3^+$ groups than that covered with 28 $-NH_3^+$ groups. The first and second peaks in the RDF plots of the former reflect CIP and SIP associations of $Cl^-$ and $-NH_3^+$ groups. Snapshots in Figure 11(e) and Figure 11(f) illustrate, respectively, CIP and SIP adsorption states of $Cl^-$ for DND–$NH_2$ with +84 charges in NaCl solution. Interestingly, we observe in Figure 11(e) that one –NH$_2$ and one $-$NH$_3^+$ compete together with 4 water molecules to share one of their partially positive charged H atoms with Cl$^-$.

The relative density of salt ions in layers perpendicular to {100} and {111} facets of DND–NH$_2$ with different charges are shown in Figure 12. An interesting feature of these plots is the existence of more pronounced peaks in $g_\perp^{\{111\}}(r_\perp)$ than in $g_\perp^{\{100\}}(r_\perp)$ for Cl$^-$ anions. It implies that there are, on average, more adsorbed Cl$^-$ onto positively charged {111} facets than {100} facets. In addition, Figure 12(a)-(b) show some adsorption of K$^+$ and Na$^+$ cations on {100} facets of the neutral DND–NH$_2$, while it is not the case for {111} facets.

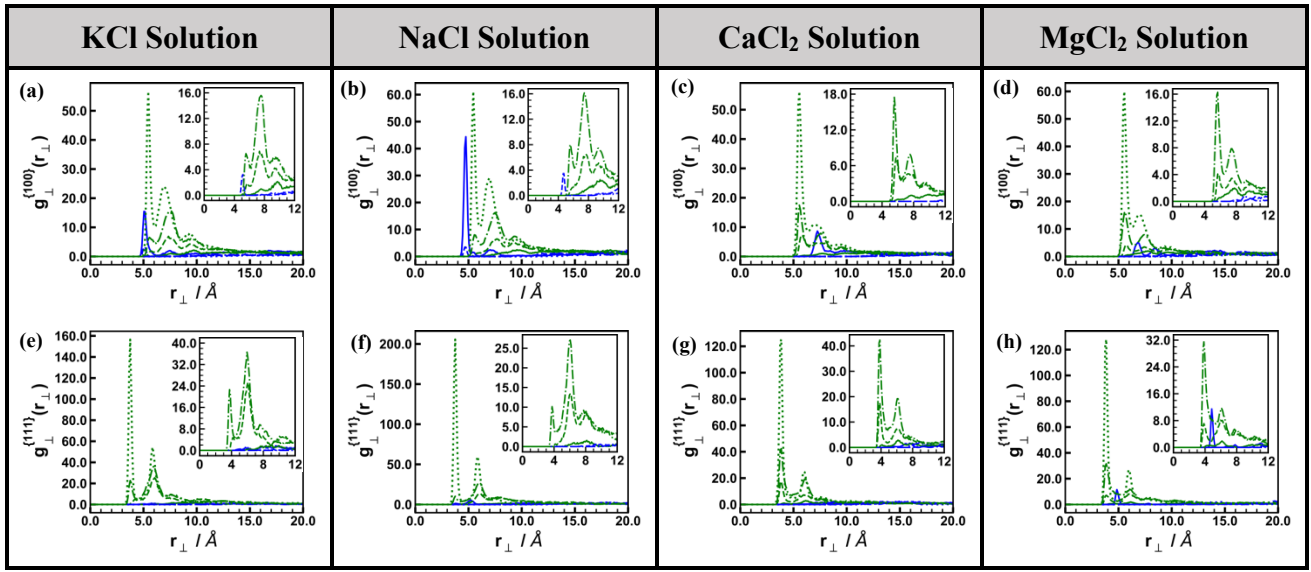

**Figure 12.** Similar to Figure 8, but for DND–NH$_2$ with 0, +28, +56, and +84 charges, which are shown as solid, dashed, dash-dotted, and dotted lines, respectively.

Last but not least, we analyze the partial RDFs of water around cations and Cl$^-$ anion, which helps us interpret results of this study better. We note two interesting features in the partial RDFs of water atoms around four cations that are shown in Figure 13(a-d). Firstly, the partial RDF for Mg$^{2+}$–OW has the most intense first peak among all cations in this study. Secondly, the first minimum of $g_{\text{Mg-OW}}(r)$ is zero, which implies a depletion region between the first and the second hydration shells of Mg$^{2+}$–OW. It is as if Mg$^{2+}$ together with its first hydration shell act as a unified moiety[100]. This interpretation corroborates the experimental observation[52,105] of the extremely slow exchange rate of water molecules between the hydration shells of Mg$^{2+}$. The first minimum of $g_{\text{Ca-OW}}(r)$ is also close to zero, but it does not lead to the abovementioned depletion region.

In Figure 13(e-h), snapshots of water molecules in the first hydration shell of cations are provided, which are obtained from our MD simulations. Cations are picked from the bulk region of the neutral DND–H solutions to ensure their hydration shells are untouched by any influence of the DND's surfaces. We can see in Figure 13(h) that Mg$^{2+}$ imposes an ordered structure on its first hydration shell, while water molecules in Figure 13(e) are quite randomly arranged around K$^+$. The hydration shell of Na$^+$, similar to that of Mg$^{2+}$, contains 6 water oxygens. However, dipole moments of water molecules (shown as dotted yellow arrows) in the former are less aligned with the line connecting the cation to water oxygen. Similar to K$^+$, Ca$^{2+}$ also contains 7 water oxygens in its first hydration shell with the distinction of more tightly and orderly packed water in the shell of the latter.

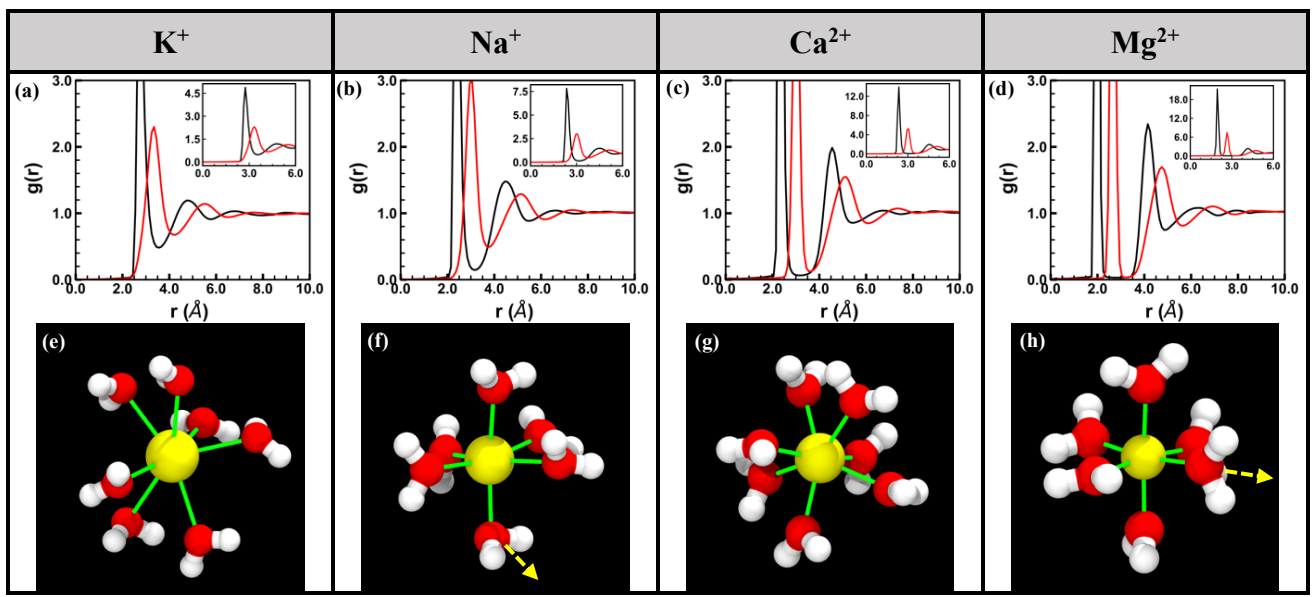

**Figure 13.** (a)-(d) Partial RDFs of water oxygen and water hydrogen around cations, (e)-(h) snapshots from MD trajectories for water molecules in the first hydration shell of cations. Black and red lines in plots correspond to water oxygen, water hydrogen, respectively. Color schemes in snapshots are the same as those in Figure 6. Dotted arrows in (f) and (h) represent the dipole moment of water.

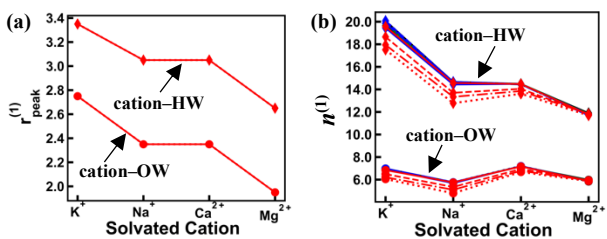

**Figure 14.** (a) The first peak position in partial RDFs of water oxygen and water hydrogen around cations, (b) the coordination number of water oxygen (OW) and hydrogen (HW) atoms in the first hydration shell of cations. The data are shown for all 52 different combinations of DND–salt solutions in this study.

In Figure 14, we have compared the partial RDF's first peak position and the coordination number in the first hydration shell of various cations, denoted as $r_{peak}^{(1)}$ and $n^{(1)}$ respectively. Values in this figure are reported for cation–chloride salt solutions of all different DNDs in this study. We observe in Figure 14(a) that $r_{peak}^{(1)}$ values for each of cation–OW and cation–HW are identical in all types of DND–salt solutions that have similar solvated cation. For instance, $r_{peak}^{(1)}$ of $K^+$–OW in KCl solution of the neutral DND–H is the same as that of DND–COOH with –84 net charges. In addition, we can order cations as $K^+ > Na^+ > Ca^{2+} > Mg^{2+}$ based on the first peak position in cation–OW's partial RDF, which trivially follows the ordering of ionic radii of cations (see Table 3). However, at a deeper level, it reflects that $Mg^{2+}$ binds water to its immediate vicinity tighter than other cations do, which arises from the significantly larger charge density of the former than that of other cations (see Table 3). The coordination numbers of water oxygen in the first hydration shell of cations in Figure 14(b) agree well with the number of oxygen atoms shown in Figure 13(e-h) as well as the reported values in the literature[33].

We note an interesting feature in Figure 14(b) for $K^+$, $Na^+$, and $Ca^{2+}$ cations in negatively charged DND–COOH solutions. Upon addition of –COO$^-$ groups to the DND's surface, the number of water molecules that hydrate $K^+$, $Na^+$, and $Ca^{2+}$ cations, i.e., $n^{(1)}$, drop. We attribute this effect to the CIP association of –COO$^-$ with those cations, which in turn become partially dehydrated as we saw earlier in Figure 6. However, we should note that coordination numbers in Figure 14(b) are averaged over all cations in the simulation box of each salt solution of DNDs. Thus, they do not reflect the true coordination number of those adsorbed cations in the CIP mode. In contrast, we do not observe in Figure 14(b) the aforementioned effect for $Mg^{2+}$. It is in accord with our earlier observation in Figure 6 that $Mg^{2+}$ cation predominantly forms SIP complex with –COO$^-$ anion.

The partial RDF of water with respect to Cl$^-$ anion is shown in Figure 15(a), which is almost identical for all aqueous salt solutions of DNDs in this study. The first peak positions in $g_{Cl-HW}(r)$ and $g_{Cl-OW}(r)$ are respectively 2.15 Å and 3.15 Å for NaCl and KCl solutions, whereas they are 2.25 and 3.25 Å for CaCl$_2$ and MgCl$_2$ solutions, respectively. We attribute the observed minor differences between chloride salts of monovalent and divalent cations to different LJ parameters that we have used for Cl$^-$ ions in the corresponding salts. However, all aforementioned values are in good agreement with reported ones in the literature[31].

The snapshot of the first hydration shell of Cl$^-$ ion, that is presented in Figure 15(b), illustrates seven water molecules that have pointed their OH bonds toward Cl$^-$. It appears as if they are competing with each other to form an HB with Cl$^-$ ions. Thus, we can expect that the HB network of water molecules in hydration layers of positively charged DND–H and DND–NH$_2$ can be disrupted upon their adsorption of Cl$^-$ anions.

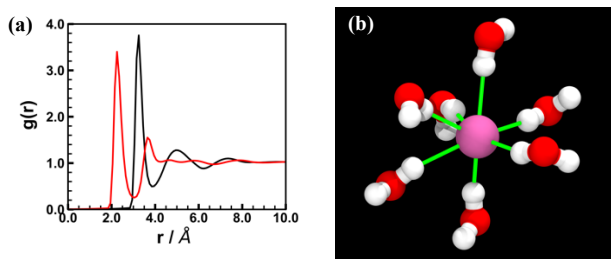

**Figure 15.** (a) partial RDFs of water oxygen and water hydrogen around Cl$^-$, (b) a snapshot from MD trajectories for water molecules in the first hydration shell of Cl$^-$ anion. Black and red lines in plots correspond to water oxygen, water hydrogen, respectively. In the snapshot, particles with red, white, and purple colors represent oxygen, hydrogen, and Cl$^-$, respectively; green lines are included just for visual aid.

### 3.2. Angular distributions

We begin this section by analyzing the 1D Angular Distribution (AD) of $\theta$ and $\beta$ angles, which are formed between the normal vector of DNDs' facets and, respectively, water dipole and water normal vectors. We have obtained ADs for water in the first hydration layers of DNDs. In addition, we have provided Distance-dependent Angular Distribution (DAD) to gain more insights into the orientations of water beyond the first hydration layers of DNDs. The DAD is a joint 2D distribution of the angle of interest and the perpendicular distance (i.e., $r_\perp$) from facets of a DND. In the ensuing analyzes, 90° for $\theta$ angle corresponds to water's dipole moment being parallel to the facet near the water molecule; a smaller or larger value than 90° means the dipole moment is oriented away or toward the facet, respectively. Furthermore, $\beta$ angles of 0° and 90° correspond to water's plane being perpendicular and parallel to the facet, respectively. Other values than 0° and 90° for $\beta$ angle indicate that water's plane is tilted with respect to the normal vector of the facet.

In Figure 16, we have shown ADs of $\theta$ and $\beta$ angles in the first hydration layer of {100} and {111} facets of DND–COOH with different surface charges in different salt solutions. A striking feature of these plots is the effect of specific adsorbed cation on both dipolar and normal orientations of water at interfaces of the negatively charged DND–COOH. In addition, we observe some differences between ADs of {100} and {111}, in particular for $\beta$ angle distributions. In what follows, we have discussed about both effects in more details.

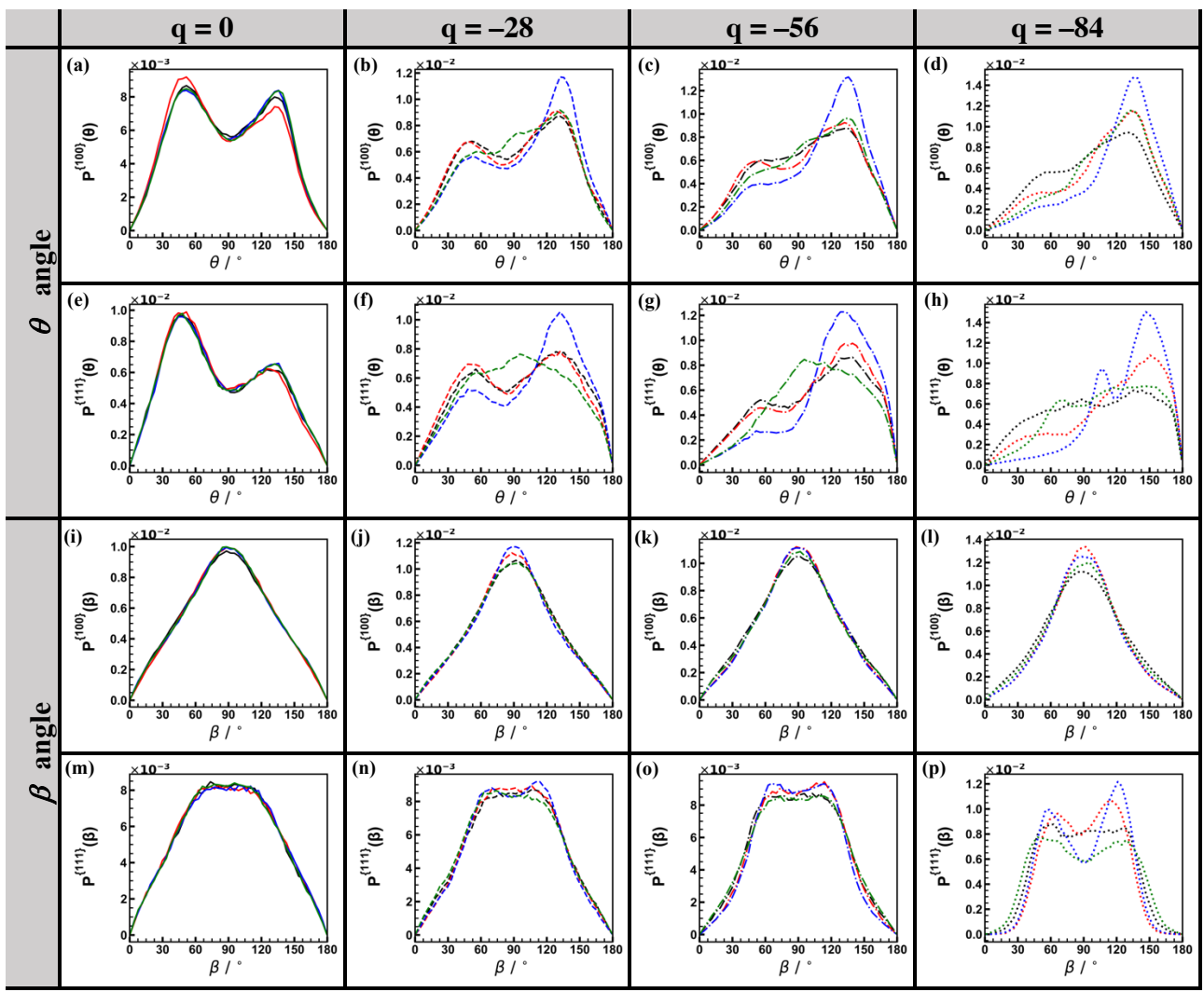

**Figure 16.** ADs of $\theta$ and $\beta$ angles in the first hydration layer of {100} and {111} facets of DND–COOH with different surface charges (q = 0, –28, –56, –84) that are solvated in four different salt solutions: KCl (red line), NaCl (black line), CaCl$_2$ (green line), and MgCl$_2$ (blue line).

As it can be seen in Figure 16(a, e), both $P^{\{100\}}(\theta)$ and $P^{\{111\}}(\theta)$ exhibit bimodal distributions for the uncharged DND–COOH. Their peak angles are at around 52° and 132°, which are averaged over values corresponding to four solutions. The first peak corresponds to water's dipole pointing away from the surface, which arises from the HB between H atom in –COOH and water oxygen. The second peak is attributed to the attraction of the water hydrogen toward either of oxygen atoms in –COOH. Thereby, water orients its dipole vector toward the surface. We note that the second peak is shallower than the first one in $P^{\{111\}}(\theta)$, which is not the case in $P^{\{100\}}(\theta)$, except for KCl solution.

The abovementioned first peak gradually diminishes in intensity, as DND–COOH acquires more negative charges. Particularly for DND–COOH with –84 charges, it either completely disappears in $MgCl_2$ solution or becomes a broad shoulder in other three salt solutions. It indicates that –COO$^-$ surface moieties on the one hand and the specific adsorbed cation on the other hand cooperatively govern the dipolar orientation of the interfacial water. The former via its O atoms act as acceptors of HBs from the interfacial water, which in turn drive the water dipole to orient toward the DND–COOH's surface. The adsorbed cation, depending on its charge density, can influence water's dipole reorientations via electrostatic interactions. In particular, $Mg^{2+}$ with a high charge density significantly restricts the reorientation of water's dipole moment.[21,30,106–108] The combined effect of –COO$^-$ and $Mg^{2+}$ appears as the sharpest peaks in $P^{\{100\}}(\theta)$ and $P^{\{111\}}(\theta)$ for the charged DND–COOH. The predominant preference of $Mg^{2+}$ for SIP over CIP association with –COO$^-$, as opposed to other cations (see Figure 6), also contributes to the aforementioned effect. Because the layer of water shared between $Mg^{2+}$ and –COO$^-$ in SIP complex can more prominently experience their cooperative effect.

Regarding $\beta$ angle distributions, we observe outstanding differences between two types of facets in Figure 16. $P^{\{100\}}(\beta)$ contains relatively sharp peaks around 90° for all surface charge values and types of salts. It implies that the most probable orientation for water's plane nearby {100} facets is to be approximately perpendicular to the facet. However, $P^{\{111\}}(\beta)$ plots exhibit distinct behaviors for different types of salts and charges on the DND. For the neutral DND–COOH in all salt solutions, $P^{\{111\}}(\beta)$ demonstrates a unimodal distribution with a broad peak. For the charged DND–COOH in $MgCl_2$ solution, $P^{\{111\}}(\beta)$ distributions display bimodal characteristics whose peaks become sharper with increasing negative charges on the DND. We observe a similar trend for KCl solution, yet with a lesser intensity. In contrast, $P^{\{111\}}(\beta)$ distributions for NaCl and $CaCl_2$ solutions of DND–COOH with various charges remain almost unimodal with a noticeable broad peak, except for q = –84. In the latter, bimodal distributions with very shallow peaks emerge. Thus, water molecules in the first hydration layer of DND–COOH with –84 net charges at {111} facets most likely tilt their planes with respect to the facet's normal vector. In this situation, the water plane's tilt angle would be most likely around 60°.

DADs for DND–COOH with –84 net charges solvated in four different salt solutions are presented in Figure 17. There are two distinct features in these distributions that are noteworthy. First, we observe that the impact of both types of facets, in particular {111} ones, on either of water's orientations goes beyond the first hydration layer. Second, contour plots associated with $MgCl_2$ solution appear to be much less diffuse than other solutions. It implies that water in hydration layers of negatively charged DND–COOH with adsorbed $Mg^{2+}$ is more neatly arranged than that with adsorbed $K^+$, $Na^+$, or $Ca^{2+}$ cations.

Angular distributions of water around DND–OH are demonstrated in Figure 18. Distributions in Figure 18(a-d) correspond to ADs in the first hydration layer, which indicate little effect of the type of the salt existing in the solution. This result is expected, as very few cations/anions are adsorbed to surfaces of DND–OH due to its overall charge neutrality in this study. However, a distinct feature of these plots is the difference between the shape of $\theta$ and also $\beta$ distributions at {100} and {111} facets. While $P^{\{111\}}(\theta)$ appears to be a bimodal distribution with two peaks at around 45º and 90º and a shoulder at around 120º, $P^{\{100\}}(\theta)$ exhibits a unimodal distribution whose broad peak at around 104º is preceded by a shoulder at around 30º. The shoulder in $P^{\{100\}}(\theta)$ and the first peak in $P^{\{111\}}(\theta)$ represent outward orientation of water dipole with respect to {100} and {111} facets. In contrast, the broad peak in $P^{\{100\}}(\theta)$ and the second peak in $P^{\{111\}}(\theta)$ as well as its shoulder correspond to water molecules orienting their dipole vectors either downward toward or parallel to respective facets.

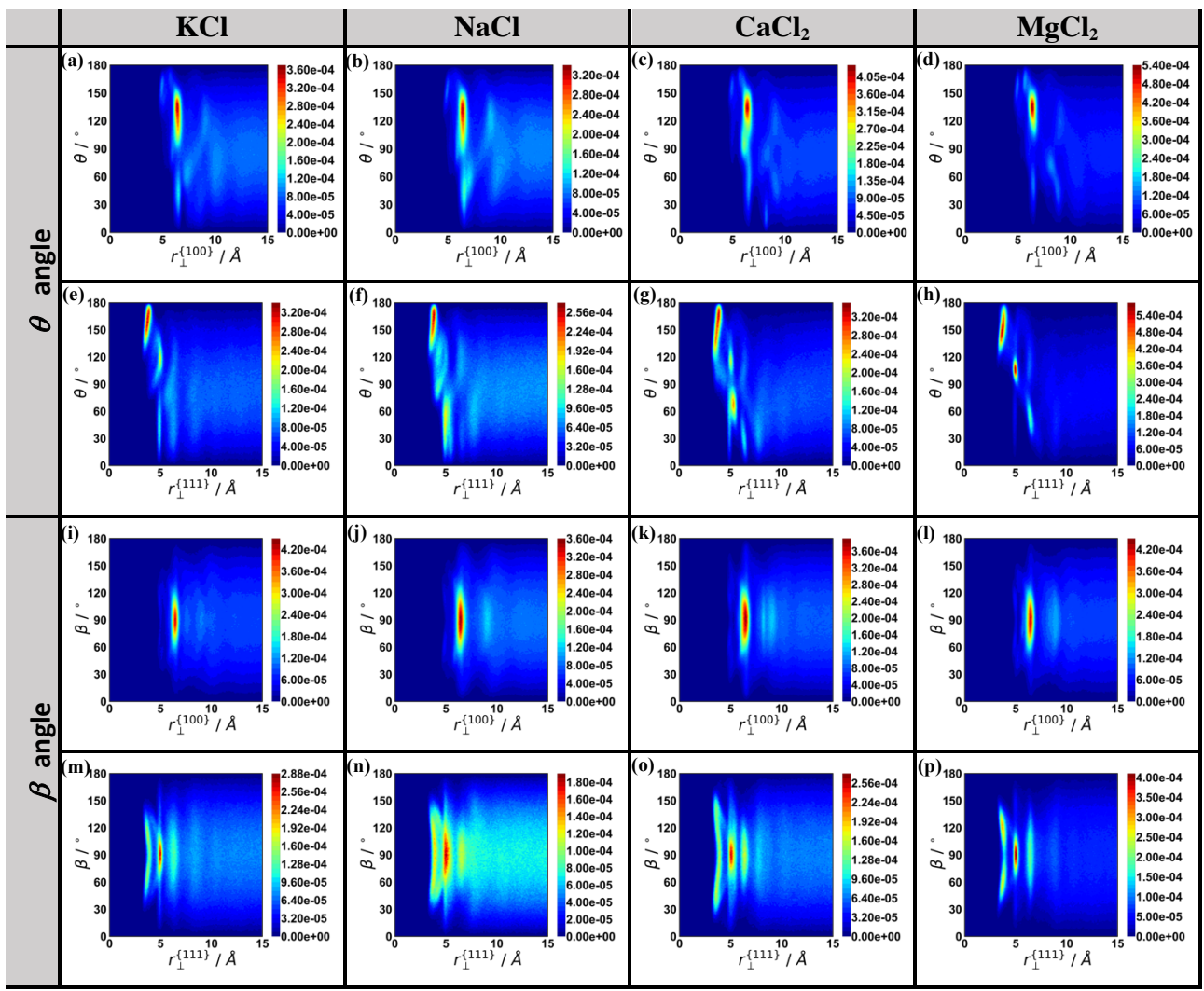

**Figure 17.** DADs for DND–COOH with –84 charges in four different salt solutions over following joint distributions: (a)-(d) $\theta - r_\perp^{\{100\}}$, (e)-(h) $\theta - r_\perp^{\{111\}}$, (i)-(l) $\beta - r_\perp^{\{100\}}$, and (m)-(p) $\beta - r_\perp^{\{111\}}$. $r_\perp^{\{100\}}$ and $r_\perp^{\{111\}}$ denote perpendicular distances from {100} and {111} facets.

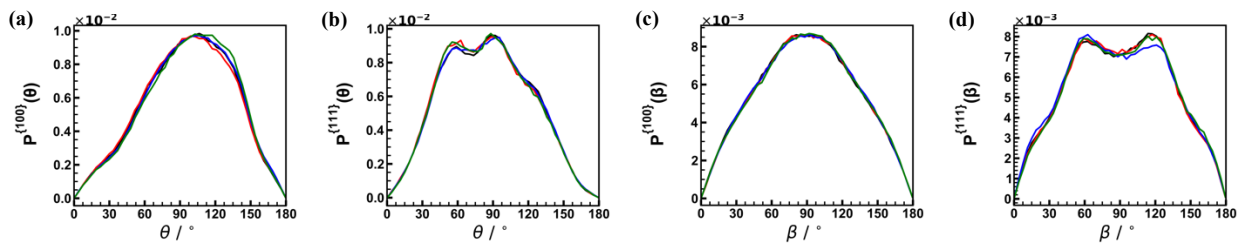

**Figure 18.** (a)-(b) ADs of $\theta$ angle in the first hydration layer of, respectively, {100} and {111} facets of DND–OH in four different salt solutions: KCl (red line), NaCl (black line), CaCl₂ (green line), and MgCl₂ (blue line), (c)-(d) the same as (a)-(b) but for $\beta$ angle.

22Regarding $\beta$ angle's distributions, the broad peak in $P^{\{100\}}(\beta)$ around 90º indicates that the interfacial water's plane is most likely either perpendicular to {100} facets or slightly tilted with respect to the facet's normal vector. Contrary to {100} facets, $\beta$ angle's distributions on {111} facets exhibit a bimodal behavior with two peaks at around 60º and 120º as well as two shoulders at around 20º and 160º. Thus, the shoulders point out the tendency of water molecules close to {111} facets to have their plane be nearly parallel to the facet. Overall, abovementioned observations lead us to infer that hydroxyls on {100} facets are most likely to donate HBs to water, whereas it is relatively equally likely for hydroxyls on {111} facets to be both HB donor to and acceptor from the interfacial water.

We now turn our attention to DND–H for which ADs of $\theta$ and $\beta$ angles in the first hydration layer at {100} and {111} facets are demonstrated in Figure 19(a-d). We can clearly see the effect of the net positive charges of DND–H on the modification of the interfacial water's orientations with respect to facets' normal vector. Since there are only minor changes in the probability distributions with respect to the type of the salt solution or the type of the facet, we only focus here on effects of the surface charges of DND–H, solvated in NaCl solution, on the angular distributions at {100} facets.

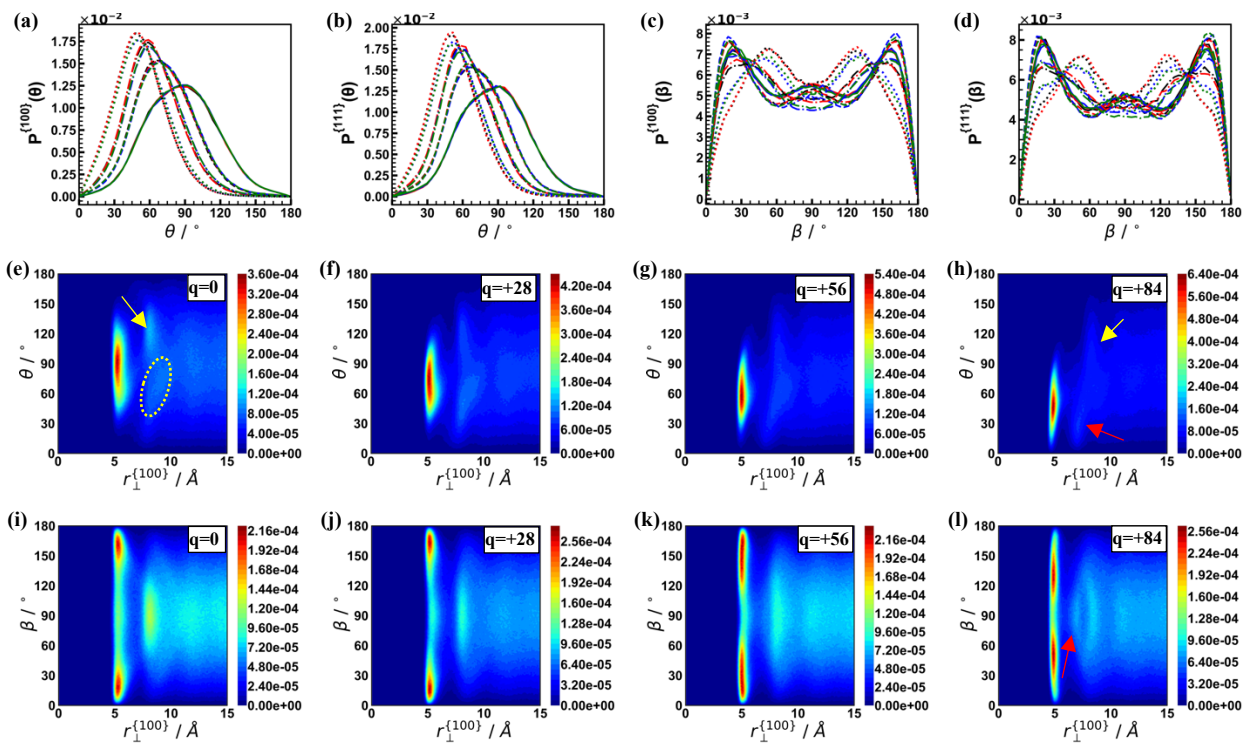

**Figure 19.** (a)-(b) ADs of $\theta$ angle in the first hydration layer of, respectively, {100} and {111} facets of DND–H in four different salt solutions: KCl (red line), NaCl (black line), CaCl$_2$ (green line), and MgCl$_2$ (blue line), (c)-(d) same as (a)-(b) but for $\beta$ angle, (e)-(h) DADs of $\theta - r_\perp^{\{100\}}$ joint distribution for NaCl solution of DND–H with four different charges (q = 0, +28, +56, +84), (i)-(l) same as (e)-(h) $\beta - r_\perp^{\{100\}}$, and (m)-(p) $\beta - r_\perp^{\{111\}}$. Solid, dashed, dash-dotted, and dotted lines in (a)-(d) represent DNDs with 0, +28, +56, and +84 charges, respectively.

At zero net charge, $P^{\{100\}}(\theta)$ displays a single broad peak ranging from 60° to 106° with a maximum at 85°. With increasing the net charge from 0 to +84 on DND–H, the aforementioned peak becomes narrower and shifted to lower angles. For DND–H with +84 net charges, $\theta$ with the highest probability is 48°. Contrary to $P^{\{100\}}(\theta)$, we observe multimodal behavior for $P^{\{100\}}(\beta)$ distributions. That is, they have three peaks for DND–H with 0 and +28 charges (with the second peak noticeably shallower and broader than the other two) and two peaks for the case of +56 and +84 charges on DND–H. Interestingly, the peaks in $P^{\{100\}}(\beta)$ for DND–H with +84 charges are closer to 90º than those of DND–H with +56 charges and also the first and third peaks of DND–H with 0 and +28 charges.

We ascribe the broad second peak around 90º in $P^{\{100\}}(\beta)$ for DND–H with 0 and +28 charges to water molecules that have at least one dangling OH (not engaged in any HB) oriented toward the associated facet. Yet, the probability of finding such water molecules nearby {100} facets is higher in the former than in the latter. In contrast, water does not have any dangling OH at {100} facets of DND–H with +56 and +84, which can be inferred from two observations: 1) the disappearance of the aforementioned peak in their $P^{\{100\}}(\beta)$ plots, 2) the relatively sharp peaks in their $P^{\{100\}}(\theta)$ distributions with peak angles at 55º and 45º, respectively. In fact, the surface H atoms on DND–H with +56 and +84 charges have higher positive partial charges than those on DND–H with +0 and +28 charges. It leads to stronger electrostatic interactions between water and surface H atoms of DND–H with +56 and +84 charges compared to DND–H with +0 and +28 charges. Thus, it appears that the sufficiently strong Coulomb potential of surfaces in DND–H with +56 and +84 charges repel the hydrogens of and instead attract the oxygen of the interfacial water. This interpretation also concurs with trends that we observed in the density plots of Figure 9.

Snapshots in Figure 20 illustrate the aforementioned different behavior of the interfacial water at one of {100} facets of DND–H with various charges. In particular, snapshots of DND–H with +56 and +84 show one adsorbed Cl⁻ anion in each case that has made the CIP association with surface H atoms. Interestingly, water molecules surrounding Cl⁻ anion, over which they compete with each other for its extra electron, have pointed one of their OH bonds away from the surface to be directed toward the anion. Thus, it appears that adsorbed Cl⁻ anions reinforce the effect of surface H atoms in pushing away hydrogen atoms of water molecules that exist in the anion's first hydration shell.

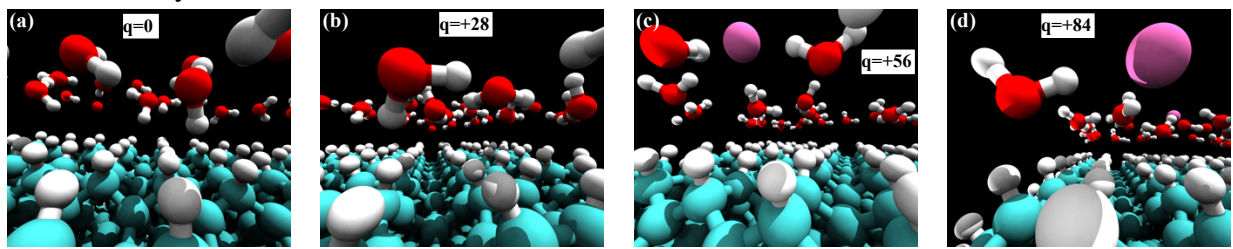

**Figure 20.** Snapshots of water molecules in the first hydration layer of DND–H with (a) 0, (b) +28, (c) +56, and (d) +84 charges. Water molecules are selected at one of {100} facets of DND–H. Particles with cyan, red, white, and purple colors represent carbon, oxygen, hydrogen, and Cl⁻, respectively.

To gain more insights into the orientations of water beyond the first hydration layer of DND–H, we investigate DADs that are shown in Figure 19(e-h) and Figure 19(i-l) for various surface charges; the former and the latter are for $\theta - r_\perp^{\{100\}}$ and $\beta - r_\perp^{\{100\}}$ joint distributions, respectively. In Figure 19(e), we have highlighted two regions with a yellow arrow and a yellow oval in the second hydration layer of {100} facets of the neutral DND–H. The former and the latter, referred to as $W_1$ and $W_2$, represent two different groups of water with their dipole moments pointed toward and away from the facet, respectively.

The orientations of $W_1$ and $W_2$ suggest that they are probably HB donors to and acceptors from water in the first hydration layer, respectively. Furthermore, higher DAD values in $\theta - r_\perp^{\{100\}}$ distributions for $W_1$ than those of $W_2$ indicate that $W_1$ has higher chances of existence than $W_2$ for the neutral DND–H. However, three trends emerge with increasing positive charges on DND–H. First, DAD values for $W_1$ diminish relative to those of $W_2$. Second, water with the characteristics of $W_1$ group appears in the second hydration layer at a longer distance from the facet, particularly for DND–H with +84 that is marked by a yellow arrow in Figure 19(h). Third, the peak corresponding to $W_2$ in DAD of $\theta - r_\perp^{\{100\}}$ for DND–H with +84 charges (pointed by a red arrow in Figure 19(h)) is closer to the facet and shifted to a smaller $\theta$ angle (ca. 30º), compared with that for the neutral DND–H. Interestingly, the last trend also manifests in DAD of $\beta - r_\perp^{\{100\}}$ for DND–H with +84 charges, where $W_2$ water has a distinct $\beta$ angle distribution than the rest of water in the second hydration layer. These trends are a direct result of disappearance of water with dangling OH nearby DND–H with +56 and +84 charges. Thus, relatively high amounts of positive charges on surfaces of DND–H can impart ordered structure in its two hydration layers.

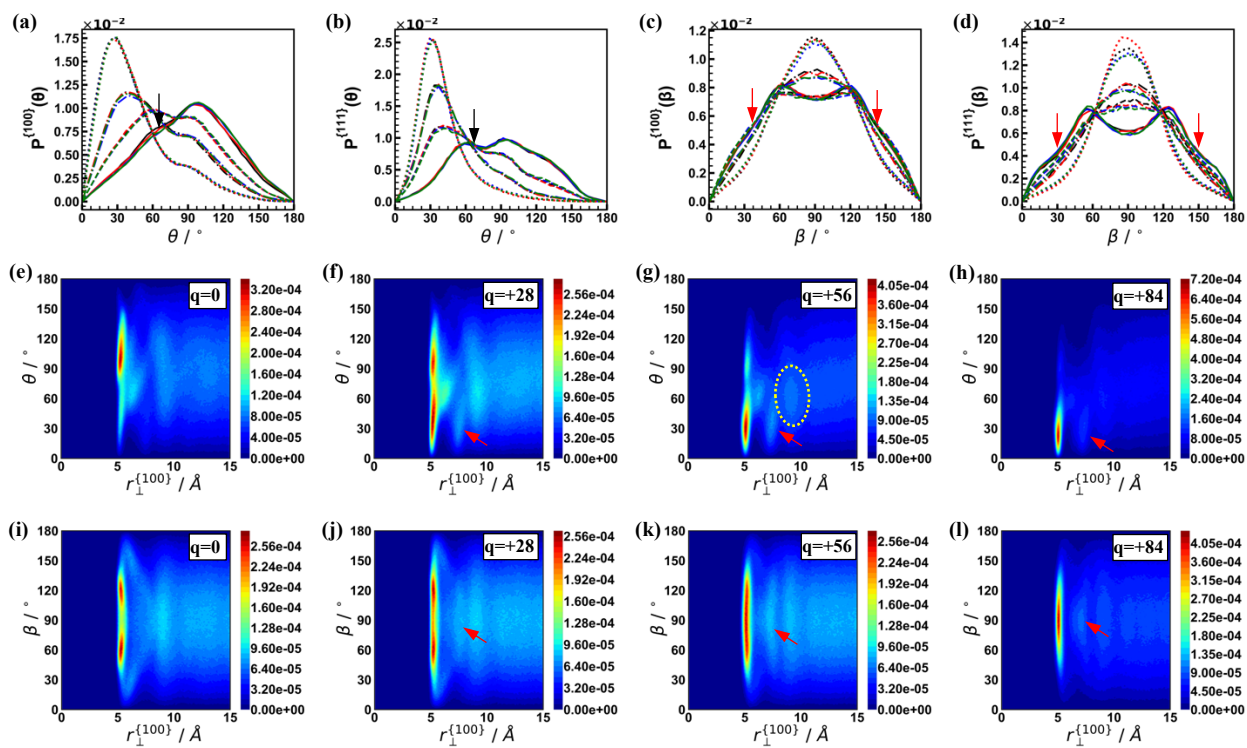

**Figure 21.** Same as Figure 19 but for DND–NH$_2$.

We move forward to discuss water orientation distributions around DND–NH$_2$ surfaces. ADs in the first hydration layer of {100} and {111} facets of this DND are shown in Figure 21(a-d), for various surface charges and salt solutions. We observe that the distributions for both $\theta$ and $\beta$ angles exhibit some similarities and differences between two types of facets. For both types of facets of the neutral DND–NH$_2$, $\theta$ and $\beta$ distributions demonstrate unimodal ($\theta_{\text{peak}} \approx 95°$) and bimodal ($\beta_{\text{peak}} \approx 63°$ and $117°$) distributions, respectively. Furthermore, the former and the latter have, respectively, one and two shoulders (pointed by black and red arrows in Figure 21), although they are more pronounced for {111} facets than those of {100} facets. Peaks of $\theta$ and $\beta$ distributions are related to each other and together they correspond to water molecules with orientations as follows: 1) one OH bond is pointed away from the facet, 2) the other OH bond is oriented toward the facet to make HB with N atom of –NH$_2$ group, and 3) the dipole moment is almost parallel to the facet. Similarly, aforementioned shoulders of these distributions are also associated with each other. They represent water molecules whose oxygens are attracted to the facet (mostly likely by H atom in –NH$_2$ group) such that their dipole moments are oriented away from the facet. The bimodal character of $\beta$ distribution comes from the fact that both OH bonds of water are equally likely to be oriented toward the facet to make HB with –NH$_2$ groups. The snapshot from the MD trajectory of the neutral DND–NH$_2$ in NaCl solution, that is shown in Figure 22(a), clearly shows these two types of preferential orientations of water.

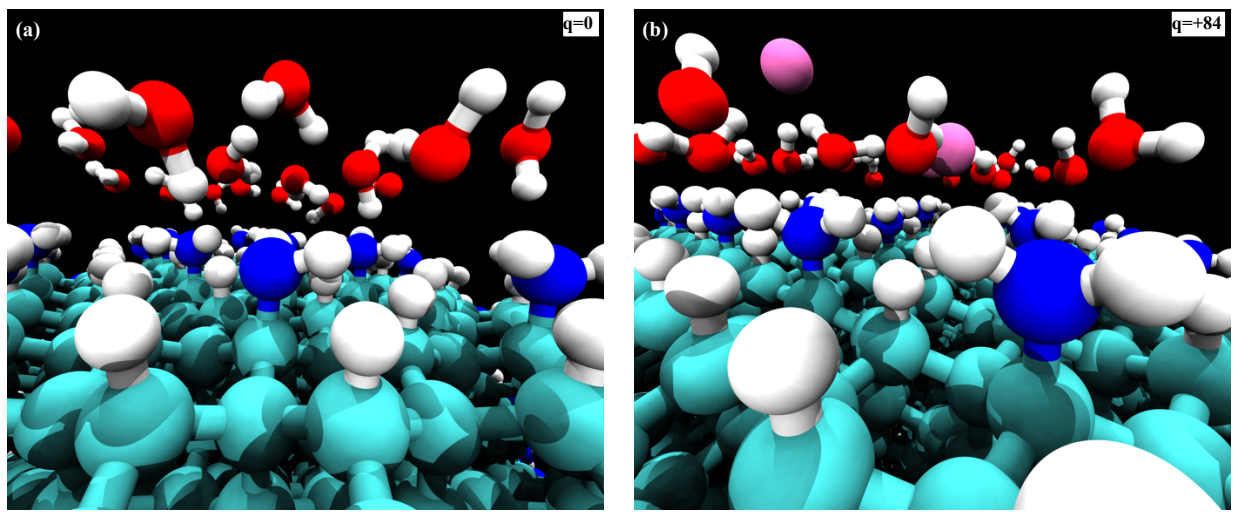

**Figure 22.** Same as Figure 20 but for DND–NH$_2$ with (a) 0 and (b) +84 charges.

The abovementioned distributions' behavior changes, as DND–NH$_2$ becomes positively charged via the protonation of some –NH$_2$ groups. With increasing positive charges on DND–NH$_2$, the major peak in $\theta$ angle distribution shifts to smaller values than 90º and $\beta$ angle distribution switches from the bimodal shape of the uncharged DND to unimodal with $\beta_{peak} \approx 90º$. Furthermore, the peak in $\theta$ angle distribution of the neutral DND–NH$_2$ appears as a shoulder in that of the charged DND–NH$_2$, whose probability values diminish with increasing positive charges on the DND. Interestingly, the shoulder disappears in $\theta$ angle distribution of {111} facets on DND–NH$_2$ with +84 charges, as opposed to that for {100} facets. In contrast, we can hardly identify any shoulder in $\beta$ angle distributions of the charged DND–NH$_2$, particularly the DND with +84 charges.

We attribute the peaks in $\theta$ and $\beta$ angle distributions to an ensemble of water molecules, where both OH bonds and hence the dipole moment of each water are driven away from the associated facet. These effects result from the relatively strong electrostatic potential of $-NH_3^+$ group that repels hydrogens of water as far as and attracts its oxygen as close as possible. These effects of $-NH_3^+$ group are more intensified in the case of DND–NH$_2$ with +84 charges, which are manifested in the relatively sharp peak of $\theta$ around 30º for both types of facets (see Figure 21(a-b)). We note that the peak positions for {100} and {111} facets are almost identical, although the peaks of the latter are sharper than the former. We note relatively sharper and more intense peaks in both $\theta$ and $\beta$ angle distributions of {111} facets than those of {100} facets, although their peak positions are almost identical. We attribute this observation to higher count of $NH_3^+$ / area for each {111} facet than that for {100} facet in our study.

DADs of $\theta - r_\perp^{\{100\}}$ and $\beta - r_\perp^{\{100\}}$ distributions for NaCl solution of DND–$NH_2$ with four different charges are presented in Figure 21(e-h) and Figure 21(i-l), respectively. We see that a new peak (specified by red arrows in Figure 21(f-h) and Figure 21(j-l)), emerges in these DADs for charged DND–$NH_2$. It represents a group of water in the second hydration layer that point their dipole moment away from the facet in an attempt to accept HBs from water in the first hydration layer. However, DND–$NH_2$ with +84 charges, compared with its less charged variants, exhibits sharper and more concentrated peak around 90° in $\beta - r_\perp^{\{100\}}$ distribution corresponding to this group of water (see Figure 21(l)). It indicates that $-NH_3^+$ groups on facets of the former impart relatively highly ordered structure in water, which extends to multiple hydration layers around DND–$NH_2$ with +84.

Last but not least, we analyze orientations of water around ions in the bulk region of solutions, far beyond the influence of DNDs' surfaces. It sheds some light onto patterns we have observed thus far in the orientational distributions of water around DNDs.

In Figure 23(a-b), ADs of $\theta$ and $\alpha$ angles for water in the first hydration shell of all cations in this study are presented. In both distributions, we find the following ordering in terms of the peak intensity:

$$K^+ < Na^+ \ll Ca^{2+} < Mg^{2+}$$

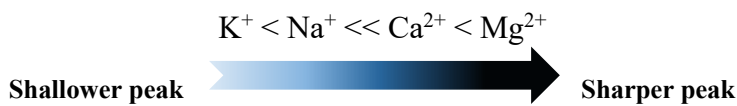

**Shallower peak**                          **Sharper peak**

Furthermore, the peak $\theta$ angles are 44°, 26°, 11°, and 11° for $K^+$, $Na^+$, $Ca^{2+}$, and $Mg^{2+}$ cations, respectively. Corresponding peak $\alpha$ angles for these cations are 74°, 70°, 60°, and 60°. Interestingly, the trends in both angles' distributions are in accord with the arrangement of water molecules in the first hydration shell of cations that we observed in Figure 13(e-h). In particular, remarkably larger peak value of $\theta$ angle for $K^+$ than that of other cations conform with chaotropic nature of the former.

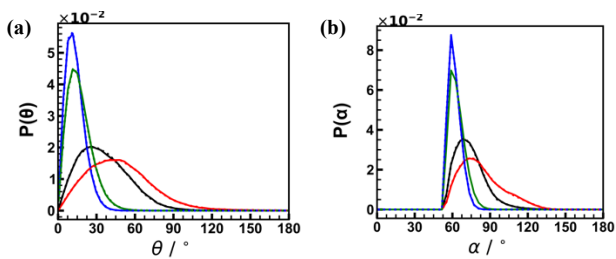

**Figure 23.** (a) ADs of $\theta$ angle in the first hydration shell of $K^+$ (red line), $Na^+$ (black line), $Ca^{2+}$ (green line), and $Mg^{2+}$ (blue line) and (b) same as (a) but for $\alpha$ angle.

DADs of $\theta - r$ and $\alpha - r$ joint distributions for water around cations are demonstrated in Figure 24(a-d) and Figure 24(e-h), respectively. Both joint probability contours in Figure 24 reveal more random arrangement of water around $K^+$ and $Na^+$ cations than that around $Ca^{+2}$ and $Mg^{2+}$ cations, even beyond their respective first hydration shells. Nevertheless, we observe an opposing behavior for $K^+$ and $Na^+$ cations. While the dipolar orientation of water looks more random around $K^+$ than that around $Na^+$, the opposite holds true for the orientation of water's OH bond. We attribute this contrasting behavior to differing charge densities of $K^+$ and $Na^+$, where the charge density of the former is almost 40% lower than that of the latter (see Table 3). Thus, the former exerts noticeably weaker electrostatic forces on the surrounding water's dipole than the latter does. As a result, the less restricted water in hydration layers of K$^+$, compared with that of Na$^+$, forms stronger hydrogen bonds with the surrounding water molecules. This interpretation also corroborates observations of Mancinelli *et al.*, where they found that Na$^+$ more effectively disrupted water structure than K$^+$ in their respective chloride solution.[92]

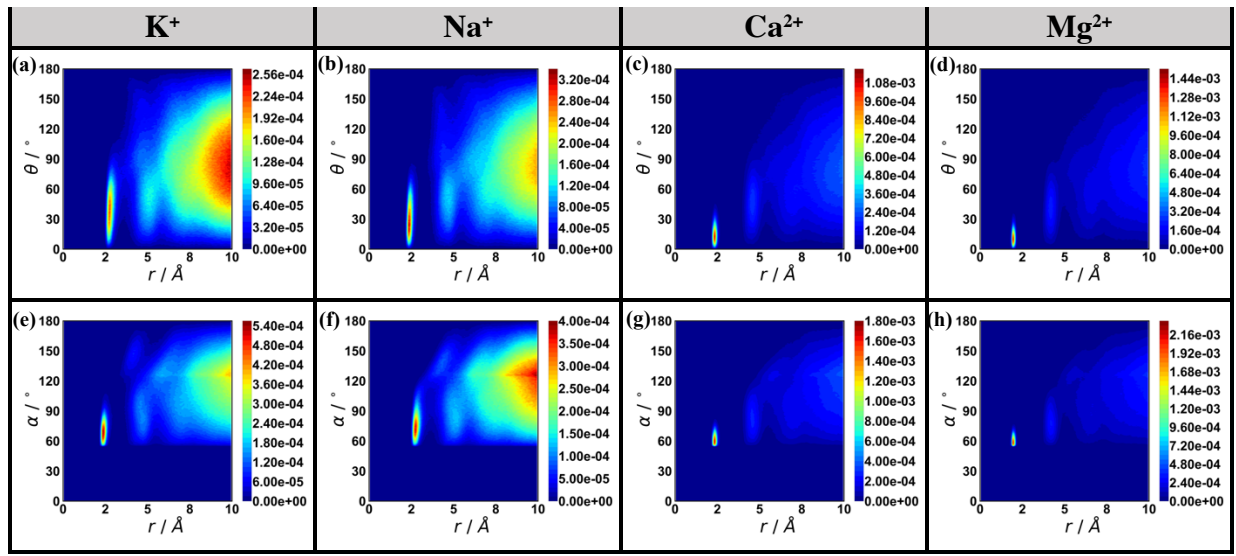

**Figure 24.** DADs of $\theta - r$ joint distribution for water around (a) K$^+$, (b) Na$^+$, (c) Ca$^{2+}$, (d) Mg$^{2+}$ cations that are solvated in the respective chloride-cation salt solution of the neutral DND–H. (e)-(h) same as (a)-(d) but for $\alpha - r$ joint distributions. $r$ is the radial distance of water oxygen from the cation.

Distribution of the $\alpha$ angle for water in the first hydration shell of Cl$^-$ anion that exist in four different salt solutions are shown in Figure 25. The peak $\alpha$ angle of 172° clearly indicates that the OH vector of water is almost aligned with the line connecting its oxygen to Cl$^-$ anion. It also agrees well with the arrangement of water around Cl$^-$ ion that we have shown in Figure 15(b).

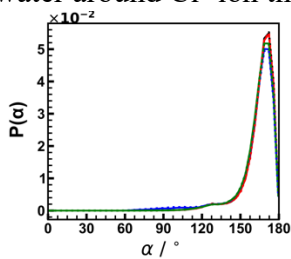

**Figure 25.** ADs of $\alpha$ angle in the first hydration shell of Cl$^-$ in four different chloride salt solutions of KCl (red line), NaCl (black line), CaCl$_2$ (green line), and MgCl$_2$ (blue line), where a neutral DND–H particle is also dissolved in each solution.

### 3.3. Residence time in hydration layers

We begin this section by analyzing the survival probability of water molecules in the first hydration shell of the dissolved anion and cations in salt solutions of the uncharged DND–H. This selection of DND helps us ensure that ions are primarily outside the hydration layers of the DND, as evidenced by density plots that we saw earlier. The Survival Correlation Function (SCF) plots related to the anion and the cations are shown in Figure 26(a) and Figure 26(b), respectively. The former compares the corresponding results for Cl⁻ anion in four different salt solutions that each contains a distinct cation.

We note in Figure 26(a) a slightly faster decay rate of SCF for hydrating water of Cl⁻ ion in CaCl$_2$ and MgCl$_2$ solutions than that corresponding to KCl and NaCl solutions, although all four curves decay to zero. We attribute the difference to the fact that LJ parameters of Cl⁻ ion in the former are different than those of the latter. In contrast, the SCF curves in Figure 26(b) for all cations, except for K$^+$, do not decay to zero within the time window of our analysis. Nonetheless, the extremely sluggish decay rate of SCF for Mg$^{2+}$ is indicative of the strikingly distinctive hydration of Mg$^{2+}$ than that of other three cations.

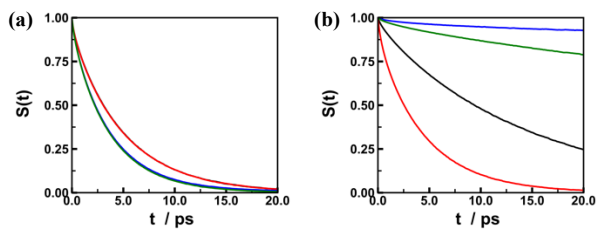

**Figure 26.** (a) SCFs of water in the first hydration shell of Cl⁻ anion in four different chloride salt solutions of KCl (red line), NaCl (black line), CaCl$_2$ (green line), and MgCl$_2$ (blue line), where each solution also includes a neutral DND–H particle. (b) same as (a) but for SCFs around the constituent cation of the solvated salt.

Except for Mg$^{2+}$ cation, we have been able to fit the tri-exponential function, which we introduced in Section 2, to the corresponding SCF curves of ions. Subsequently, the numerical integration of the fitted SCFs has given us the mean residence times for water in the first hydration shell of ions. The results are presented in Table 4, which are in good agreement with reported values in the literature. However, it is almost infeasible to obtain the same quantity for Mg$^{2+}$ from MD simulations with the existing computational power. Because the exchange rate of water between the first and second hydration layers of Mg$^{2+}$ cation is extremely slow.[83,101,109] Indeed, NMR measurements have revealed the mean residence time on the microsecond timescale for water in the first hydration shell of Mg$^{2+}$ cation.[93,96] In contrast, as we can see in Table 4, the corresponding values for all other ions are on the picosecond timescale.

**Table 4.** The mean residence time of water in the first hydration shell of ions ($\tau_{res}^{(1)}$), obtained from our own calculations and from available data in the literature.

| Ion | $\tau_{res}^{(1)}$ / ps | |
|---|---|---|
| | Our Calc. | Literature |
| **Cl⁻ (KCl solution)** | 4.6 | – |
| **Cl⁻ (NaCl solution)** | 4.6 | 4.5 [a], 4.0 [b] |
| **Cl⁻ (CaCl$_2$ solution)** | 3.5 | – |
| **Cl⁻ (MgCl$_2$ solution)** | 3.6 | – |
| **K$^+$** | 4.1 | 4.8 [a] |
| **Na$^+$** | 14.1 | 9.9 [a], 20.0 [b] |
| **Ca$^{2+}$** | 110.6 | ~100 [c] |
| **Mg$^{2+}$** | – | $2 \times 10^6$ [c] |

[a] Ref.[78], [b] Ref.[110], [c] Ref.[93,94,96,97,105].

Interestingly, we observe that the mean residence time of water in the first hydration shell of ions follows the order below, which is commensurate with their charge densities and thereby their structure-breaking/structure-making tendencies:

$$Cl^- \approx K^+ < Na^+ \ll Ca^{2+} \lll Mg^{2+}$$

Low charge density →  High charge density

The abovementioned order is also in accordance with cations' ADHN values presented in Table 3. In addition, the significantly large mean residence time for $Mg^{2+}$ explains the depletion region we noted for the partial RDF of Mg-OW in Figure 13(d).

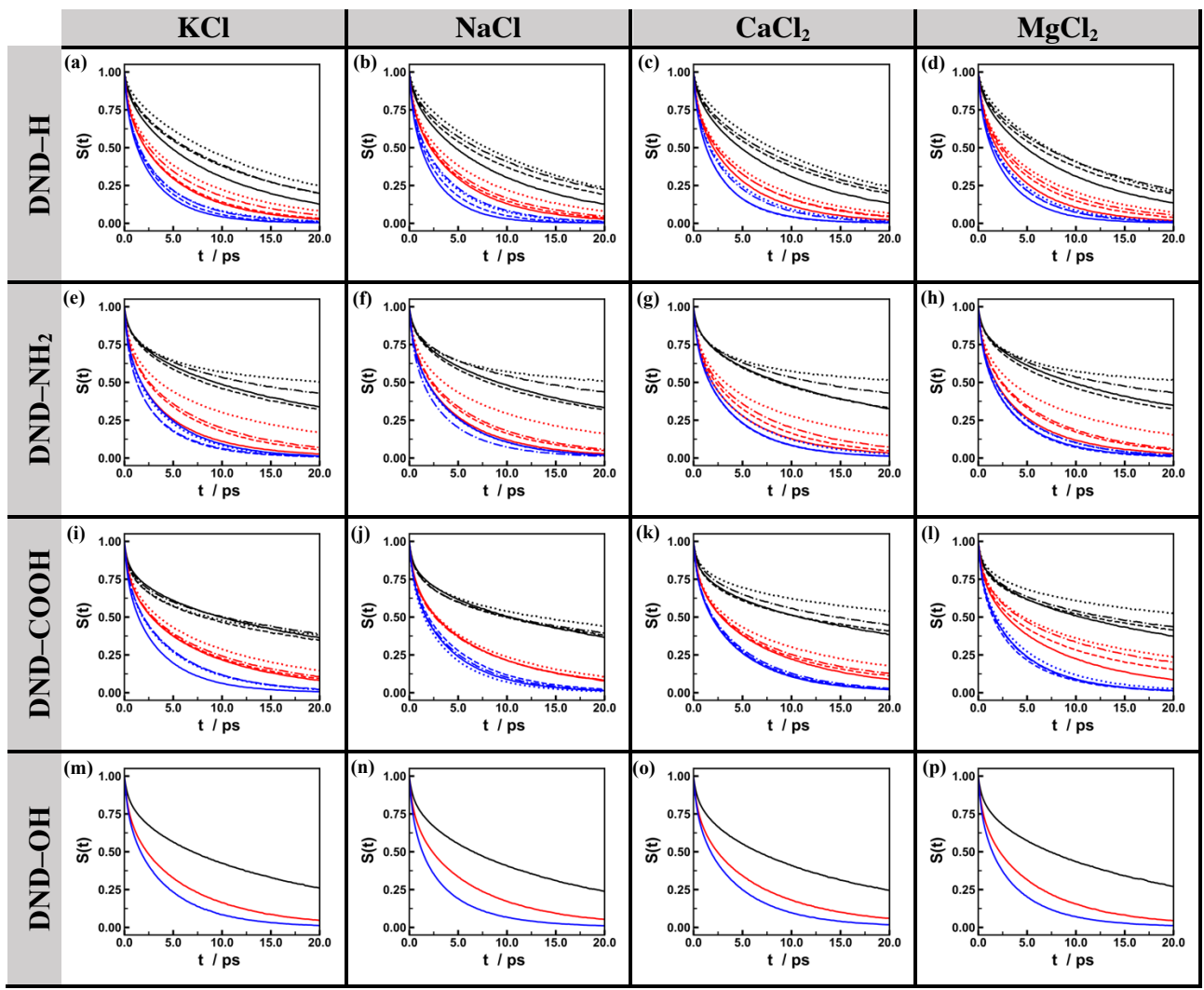

**Figure 27.** SCFs of water in the first (black line), second (red line), and third (blue line) hydration layers of different DNDs with various surface charges that are solvated in four different salt solutions. Solid, dashed, dash-dotted, and dotted lines correspond to 0, 28, 56, and 84 absolute charges on DNDs, respectively. DND–H and DND–NH$_2$ assume any of these charges with positive sign, so does DND–COOH but with negative sign. But, DND–OH only exists as neutral in our study.

We now turn our attention to the SCFs of water in each of the three hydration layers of DNDs, which are shown in Figure 27. The SCF in a particular hydration layer of a DND (e.g., the first hydration layer) has been obtained by taking the average of the SCF in the same hydration layer corresponding to 14 facets of the DND. Then, we have taken another average over results obtained from five distinct MD runs to attain better statistics.

At least, two interesting patterns in Figure 27 are worth noting. Firstly, the SCF of the third hydration layer of all DNDs decays to zero quicker than that of other hydration layers. In other words, after "$t$" time being elapsed from the arrival of water in the 3$^{rd}$ hydration layer of any DND, its survival probability rapidly drops to zero. It implies that the dynamics of water is much less constrained in the third hydration layer and thereby it can be very quickly exchanged with either the bulk water or water in the second hydration layer. Secondly, with some exceptions, the decay rate of the SCFs in the first and second hydration layers of charged DNDs appears to be slower compared with their own uncharged counterparts. It suggests that water experiences some constraints as it moves through those regions. We have discussed these constraints below with the help of the mean residence time of water in these two regions, i.e., $\tau_{res}^{(1)}$ and $\tau_{res}^{(2)}$, which are shown in Figure 28. The tabulated values for $\tau_{res}^{(1)}$, $\tau_{res}^{(2)}$, and $\tau_{res}^{(3)}$ are presented in the Supplementary Information.

We can clearly see the "specific ion effects" for the negatively charged DND–COOH in Figure 28, where $\tau_{res}^{(1)}$ values vary as a function of the adsorbed cation's nature that follows the well-known Hofmeister series[30], namely, K$^+$ < Na$^+$ < Ca$^{2+}$ < Mg$^{2+}$. With the exception of the weakly hydrated K$^+$ cation, this effect on $\tau_{res}^{(1)}$ becomes more pronounced as DND–COOH acquires more negative charges and hence adsorbs higher concentration of cations.

One might ask what mechanism at the molecular level could explain the "specific cation effect" on the exchange dynamics of water from the interface of the negatively charged DND–COOH? To answer this question, we borrow the concept of the cation-anion cooperative hydration from the literature. It states that strongly hydrated cations and anions can cooperatively slow down the orientational dynamics of water that is shared between them.[83] For instance, Pastorczak *et al.* observed that the addition of (HCOO$^-$, Na$^+$) and (HCOO$^-$, Li$^+$) salts to water significantly retarded its reorientational dynamics, which became more pronounced at higher concentrations.[100] In contrast, the aqueous solution of (HCOO$^-$, K$^+$), (I$^-$, Na$^+$), or (I$^-$, Li$^+$) did not exhibit any noticeable difference in rotational dynamics of water with respect to the pure water. In their study, HCOO$^-$, Li$^+$ and, Na$^+$ are strongly hydrated ions, whereas K$^+$ and I$^-$ are weakly hydrated ones.

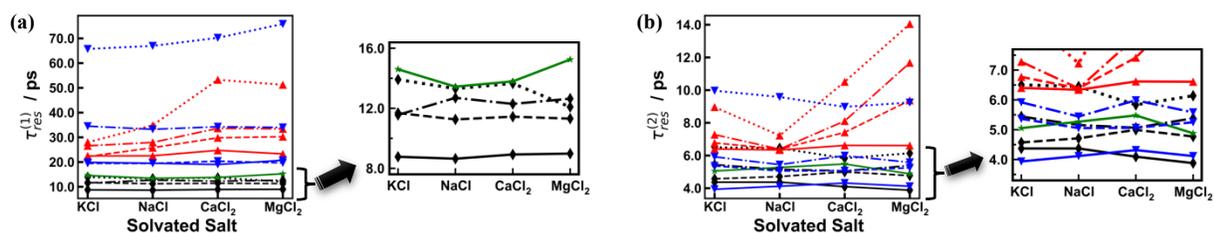

**Figure 28.** Mean residence time of water in (a) the first and (b) the second hydration layers of different DNDs with various surface charges (differentiated by different line styles). Black, blue, red, and green lines represent DND–H, DND–NH$_2$, DND–COOH, and DND–OH particles, respectively. Line styles are the same as those in Figure 27.

The basic idea is that the strongly hydrated cation locks the rotation of water's dipole moment, while the strongly hydrated anion significantly constrains the reorientation of OH bond of water. This combined effect is non-additive in the sense that the cooperative slowdown is more significant than the effect of isolated ions on the dynamics of their hydration shells.[83] Furthermore, the magnitude and the spatial extent of the cooperative retardation is proportional to the surface charge density and the concentration of ions.[83,101] The surface charge density of ions also determines their water affinity and subsequently their ion pairing propensity toward other oppositely charged ions. Prior studies suggest that strongly hydrated ions that form more SIP and even 2SIP than CIP complexations, can more significantly slow down the reorientational dynamics of water.[100,102]

The linkage between reorientational dynamics of water and its exchange dynamics in hydration layers is hidden in the rearrangements of the HB network. A water molecule cannot move around without breaking HBs with its neighbors. In turn, the latter would not be possible unless the water molecule and its neighbors rotate in order to find new HB partners.[103,104] In a relatively high concentrated solution, high charge density cations and anions such as $Mg^{2+}$ and $SO_4^{2-}$ can cooperatively lock in the HB network in multiple directions.[83] Thus, a rigid HB network induced by cooperative slowdown of reorientational dynamics of water can also lead to slowing the exchange dynamics of water in hydration layers.

We can also observe the specific cation effect on the residence time of water in the second hydration layer of the charged DND–COOH. That is, $\tau_{res}^{(2)}$ appears to depend on the nature of the adsorbed cation and increases in the sequence $Na^+ < K^+ < Ca^{2+} < Mg^{2+}$. However, there are two main differences between the first and the second hydration layers in terms of the specific cation effects. Firstly, contrary to the first hydration layer, $K^+$ cation is associated with slower exchange dynamics compared with $Na^+$. In particular, $\tau_{res}^{(2)}$ for DND–COOH with –84 charges, that has adsorbed $K^+$ cations, is 29% larger than the corresponding value for the adsorbed $Na^+$. It is quite counterintuitive as one might expect the opposite association, since the latter has a higher water affinity than the former. Secondly, $\tau_{res}^{(2)}$ values are smaller than $\tau_{res}^{(1)}$ values by approximately 4-5 orders of magnitude. These observations together also provide additional support for the cooperative slowdown mechanism, to which we ascribed the specific cation effect on $\tau_{res}^{(1)}$. In particular, the different effects of $Na^+$ on the exchange dynamics of water in the first and second hydration layers corroborate the experimental observations of Ref.[100].

We attribute the aforementioned difference between $K^+$ and $Na^+$ to two factors, which have their origin in different water affinities. First, as we observed earlier, $Na^+$ has higher propensity to form CIP associations with $–COO^-$ surface groups than $K^+$. Thus, there is a higher concentration of $K^+$ in the second hydration layer than $Na^+$. Second, $K^+$ cation, as opposed to $Na^+$, is weakly hydrated and hence water in its labile hydration shells has more freedom to make HBs with surrounding molecules. It is also supported by more random orientations of water OH bonds around $Na^+$ than those around $K^+$ that we observed in Figure 24(e-f). Thus, the net effect of these factors is that $K^+$ induces a stronger HB network than $Na^+$ in the second hydration layer of DND–COOH. Thereby, it leads to slower exchange dynamics of water associated with the former than the latter.

The reason that divalent cations give rise to higher values for $\tau_{res}^{(2)}$ than the monovalent ones can be explained by the highly kosmotropic (structure-making) nature of the former. In other words, $Mg^{2+}$ and $Ca^{2+}$ cations can create ordered structures in water well beyond their first hydration shells, which thereby lend themselves to more rigid HBs network in their surroundings. Particularly, six water molecules, which are tightly bound to the first hydration layer of $Mg^{2+}$, altogether donate 12 strong HBs to water molecules in the second hydration shell of $Mg^{2+}$ cation.[52,105] However, the structure-making effects of $Na^+$ cation, which is also considered a strongly hydrated ion, is limited to its first hydration shell.

We proceed to analyze the residence time of water in the first two hydration layers of DND–NH$_2$. Strikingly, compared with all other DNDs solvated in different solutions, water exhibits the slowest exchange dynamics in the first hydration layer of DND–NH$_2$ with positive charges of 56 or higher. However, at lower surface charges, the DND–COOH give rise to higher values for $\tau_{res}^{(1)}$. We attribute the aforementioned slow exchange dynamics to strong HBs that are donated by NH$_3^+$ surface groups to the interfacial water. However, their collective effect only becomes pronounced after being grafted onto surfaces of DND–NH$_2$ in sufficiently large amounts (here, at or greater than 56 groups in total). It is in accordance with AD patterns of the DND–NH$_2$ that we saw in Figure 21. That is, the interfacial water of DND–NH$_2$ with positive charges of 56 or higher predominantly orients its oxygen toward and its OH bonds away from the DND's facets.

The aforementioned preferential orientation also explains why DND–NH$_2$ with +84 charges gives rise to higher values for $\tau_{res}^{(2)}$ in KCl or NaCl solutions compared with DND–COOH with –84 charges. In other words, water in the first hydration layer of the former donates relatively strong HBs to water molecules in the second hydration layer. Thus, the mobility of water in the second hydration layer of DND–NH$_2$ with +84 charges is more restricted than that of DND–COOH with –84 charges, provided that both are solvated in either KCl or NaCl solutions. In contrast, in CaCl$_2$ and MgCl$_2$ solutions, the charged DND–COOH dominates the charged DND–NH$_2$, due to the ordered HB network induced by strongly hydrated adsorbed Mg$^{2+}$ or Ca$^{2+}$ in the second hydration layer of DND–COOH with –84 charges.

The mean residence time of water in the hydration layers of DND–H also reveals interesting patterns. Water in the first hydration layer of the neutral DND–H has the lowest mean residence times compared with other DNDs, irrespective of the type of the solvated salt. We attribute this effect to the hydrophobicity of the neutral DND–H, which is also supported by the AD plots shown in Figure 19. However, as the number of positive charges on surfaces of DND–H increases, so does the corresponding value of $\tau_{res}^{(1)}$. Interestingly, $\tau_{res}^{(1)}$ values of DND–H with +84 are very close to those of DND–OH in all salt solutions, except MgCl$_2$ solution. It suggests that positively charged DND–H modifies the HB network of its interfacial water with respect to its neutral counterpart. Similar to $\tau_{res}^{(1)}$, we also observe an association between positive charges of DND–H and the corresponding values of $\tau_{res}^{(2)}$, as shown in Figure 28(b). Furthermore, it appears in Figure 27 that the SCFs in the third hydration layer of DND–H with higher positive charges decay at lower rates. These two observations suggest that the impact of the aforementioned modified HB network at the interface of positively charged DND–H also extends to its second and third hydration layers. Although these interpretations qualitatively corroborate previously reported experimental results[33], a quantitative investigation of such HB network's modification is required.

## 4. Conclusion

We have carried out MD simulations to study the hydration layers formed around cuboctahedral DNDs with various surface chemistries that are solvated in four different chloride salt solutions. In general, we have identified three hydration layers around facets of DNDs by exploiting normalized density plots perpendicular to those facets. These layers are characterized by, respectively, a sharp peak followed by a shallower peak and finally a small bump in aforementioned plots. The layering in the interfacial water disappears at around 1 nm from the outermost atom layer on facets of DNDs. It is in good agreement with what have been reported in the literature for DNDs[31] and other faceted particles[21,30,106–108].

We note distinct structural and dynamic features between the first and other two hydration layers. We have summarized these features below, which are a manifest of DNDs' surface functional groups and adsorbed ions on polar and/or charged surfaces.

1) The mean residence time of water in the first hydration layer of DNDs is substantially higher than that of water in other layers. It indicates that there is a region close to DND's surfaces, where the hydrating water behaves differently compared with the rest of hydration layers. This is in accordance with the existence of an inert region around DND, where water remains intact upon exposures to freezing/melting experiments.[31,111] However, they had not considered effects of solvated ions and different DND's functional groups on this inert region. Indeed, we have found that $Mg^{2+}$ and $Ca^{2+}$ cations around negatively charged DND–COOHs can extend this region even further away from DND's surfaces. Furthermore, water in the first hydration layer of DND–NH$_2$ with +84 net charge has the longest mean residence time compared with all other studied DND solutions. We attribute this effect to strong HBs between water and $NH_3^+$ groups on charged DND–NH$_2$. However, at lower net absolute charge on DNDs, water on average spend the longest time in the first hydration layer of DND–COOH solvated in MgCl$_2$. In addition, the shorter mean residence time of water in hydration layers around DND–H than that of the oxidized DNDs corroborates thermal analysis experiments reported elsewhere[32].

2) We report different behavior in adsorption of cations onto surfaces of DNDs. We have found $Mg^{2+} \ll K^+ < Ca^{2+} < Na^+$ ordering for tendency of cations to form Contact Ion-Pair (CIP) with –COO$^-$ anion on surfaces of the negatively charged DND–COOH. Indeed, $Mg^{2+}$ predominantly forms Solvent-shared Ion-Pair (SIP) with –COO$^-$ anion, although we have observed some few CIP association between them only at high concentrations of –COO$^-$ on DND–COOH (i.e., 84 –COO$^-$ / DND). Other three cations also form SIP associations with –COO$^-$, yet with lesser proportion to their CIP complexations. Na$^+$ and K$^+$ cations, but not Ca$^{2+}$ and Mg$^{2+}$, also form CIP associations with –OH, –COOH, –NH$_2$ polar groups of, respectively, DND–OH, neutral DND–COOH, and neutral DND–NH$_2$. However, divalent cations show some weak tendencies to form SIP complexations with those polar groups. The above-mentioned trends for cation adsorption on –COO$^-$ anion agree well with prior experimental and MD simulation studies[99,112–114]. In addition, we have found that the law of matching water affinity, which is advocated by Collins[52,115,116], explains very well all but two of our observations of cation adsorption behavior. In particular, it fails to explain why Mg$^{2+}$ forms few CIP associations with –COO$^-$ anion in the case of DND–COOH with –84 charges. In fact, Mg$^{2+}$ strongly binds water to its immediate vicinity, owing to its high charge density. Thus, the CIP association with –COO$^-$ requires significant amounts of energy to partially dehydrate Mg$^{2+}$ cation[117]. We hypothesize that this energy is provided by sufficiently high concentrations of –COO$^-$ on DND–COOH, which in turn is compensated by the entropy gain resulted from more adsorption of Mg$^{2+}$ cations. We rationalize the entropy gain based on our observations that adsorbed Mg$^{2+}$ cations induce a highly ordered structure in hydration layers of the negatively charged DND–COOH. Since the fully hydrated Mg$^{2+}$ appears as a large sphere, it would be difficult to accommodate high concentrations of them on surfaces of DND–COOH terminated with relatively high concentrations of –COO$^-$. Thus, more space created by the CIP association of Mg$^{2+}$–COO$^-$ facilitates more Mg$^{2+}$ adsorption. Thereby, they can impart more order in hydration layers of DND–COOH.

3) Charged polar groups on DNDs significantly modify the preferred orientation of the interfacial water. Replacing 26% of the initial 320 –NH$_2$ groups on DND–NH$_2$ by $NH_3^+$ group makes the interfacial water shift its dipole moment from being pointed toward the surface to be oriented away from it. It implies that this surface modification converts the interfacial water from an HB donor to DND–$NH_2$ to an HB acceptor from it. We have observed similar behavior, but in opposite direction, for the interfacial water of DND–COOH by substituting 26% of the initial 320 –COOH groups with –$COO^-$ group. Furthermore, we have observed a cooperation between –$COO^-$ group and the adsorbed cation in modulating preferred orientation of the interfacial water. The former drives the water's dipole moment to point toward DND–COOH's surface to receive a HB from water. The latter can influence water's dipole reorientations via electrostatic interactions whose strength depends on the cation's charge density. In particular, the high charge density $Mg^{2+}$ cation significantly restricts the reorientation of water's dipole moment[83,101,109]. The combined effect of –$COO^-$ and $Mg^{2+}$ leads to the highest structured water around the charged DND–COOH, compared with the effect of other adsorbed cations. The more prominent cooperative effect of $Mg^{2+}$ and –$COO^-$ is also derived from the predominant preference of $Mg^{2+}$ for SIP over CIP association with –$COO^-$, as opposed to other cations. We also attribute the specific cation effect on the mean residence time of water in hydration layers of charged DND–COOH to the aforementioned cooperative hydration. We have presented a detailed discussion of this effect in Section 3.3.

4) We report noticeable modifications in the hydration layers around the hydrogenated DNDs (DND–H), as they acquire more positive charges. In particular, there exist noticeable amounts of water with dangling OH (i.e., not involved in any hydrogen bonding) at the interface with the neutral DND–H. They orient one or both of their OH bonds toward a nearby facet of the DND–H. However, they less frequently appear in the case of DND–H with +28 charges and completely disappear at the interface with DND–H having +56 and +84 charges. Thus, higher positive partial charges of surface H atoms in the latter attract oxygen of the interfacial water and push away its hydrogen atoms. Furthermore, DND–H with +84 charges imparts more order in the arrangement of water in its second hydration layer, compared with DND–H having +28 and +56 charges. The aforementioned orientation of water nearby DND–H with +84 charges corroborates results of a recent experimental study[35], they are in sharp contrast with what Petit *et al.* reported earlier[33]. DFT calculations performed by Manelli *et al.* on adsorption of water onto bare and hydrogenated {100} diamond surfaces help us resolve this apparent discrepancy.[59] Their findings suggest that water orients both of its OH bonds, and hence its dipole moment, away from the fully hydrogenated surface. In contrast, it points one of its OH bonds downward toward the bare as well as partially hydrogenated surface that contains carbon with dangling bonds. Thus, the former concurs with our aforementioned results as well as those of Ref.[35], while the latter corroborates orientations reported in Ref.[33].

**References**

1. Lanin, S. N., Platonova, S. A., Vinogradov, A. E., Lanina, S. & Nesterenko, P. N. Regularities of adsorption of water-soluble vitamins on the surface of microdispersed sintered detonation nanodiamond. *Adsorption* **24**, 637–645 (2018).
2. Huang, H., Pierstorff, E., Osawa, E. & Ho, D. Active nanodiamond hydrogels for chemotherapeutic delivery. *Nano Lett.* **7**, 3305–3314 (2007).

3.  Li, J. *et al.* Nanodiamonds as intracellular transporters of chemotherapeutic drug. *Biomaterials* **31**, 8410–8418 (2010).
4.  Perevedentseva, E., Lin, Y. C., Jani, M. & Cheng, C. L. Biomedical applications of nanodiamonds in imaging and therapy. *Nanomedicine* vol. 8 2041–2060 (2013).
5.  Lin, Y. W. *et al.* Co-delivery of paclitaxel and cetuximab by nanodiamond enhances mitotic catastrophe and tumor inhibition. *Sci. Rep.* **7**, (2017).
6.  Shimkunas, R. A. *et al.* Nanodiamond–insulin complexes as pH-dependent protein delivery vehicles. *Biomaterials* **30**, 5720–5728 (2009).
7.  Xing, Y. & Dai, L. Nanodiamonds for nanomedicine. *Nanomedicine* vol. 4 207–218 (2009).
8.  Reineck, P. *et al.* Visible to near-IR fluorescence from single-digit detonation nanodiamonds: excitation wavelength and pH dependence. *Sci. Rep.* **8**, (2018).
9.  Nunn, N. *et al.* Fluorescent single-digit detonation nanodiamond for biomedical applications. *Methods Appl. Fluoresc.* **6**, (2018).
10. Nesterenko, P. N., Fedyanina, O. N., Volgin, Y. V. & Jones, P. Ion chromatographic investigation of the ion-exchange properties of microdisperse sintered nanodiamonds. *J. Chromatogr. A* **1155**, 2–7 (2007).
11. Sakurai, H. *et al.* Adsorption Characteristics of a Nanodiamond for Oxoacid Anions and Their Application to the Selective Preconcentration of Tungstate in Water Samples. *Anal. Sci.* **22**, (2006).
12. Khanna, P. *et al.* Use of nanocrystalline diamond for microfluidic lab-on-a-chip. *Diam. Relat. Mater.* **15**, 2073–2077 (2006).
13. Peristyy, A. A., Fedyanina, O. N., Paull, B. & Nesterenko, P. N. Diamond based adsorbents and their application in chromatography. *Journal of Chromatography A* vol. 1357 68–86 (2014).
14. Xue, Z., Vinci, J. C. & Colón, L. A. Nanodiamond-Decorated Silica Spheres as a Chromatographic Material. *ACS Appl. Mater. Interfaces* **8**, 4149–4157 (2016).
15. Brenner, D. *et al.* Nanodiamond-based Nanolubricants: Experiment and Modeling. in *Mater. Res. Soc. Symp. Proc* 1703 (2014). doi:10.1557/opl.2014.
16. Liu, Z. *et al.* Tribological properties of nanodiamonds in aqueous suspensions: Effect of the surface charge. *RSC Adv.* **5**, 78933–78940 (2015).
17. Ivanov, M. & Shenderova, O. Nanodiamond-based nanolubricants for motor oils. *Curr Opin Solid State Mater Sci* **21**, 17–24 (2017).
18. Edgington, R. *et al.* Functionalisation of Detonation Nanodiamond for Monodispersed, Soluble DNA-Nanodiamond Conjugates Using Mixed Silane Bead-Assisted Sonication Disintegration. *Sci. Rep.* **8**, (2018).
19. Tiainen, T., Myllymaki, T., Hatanpaa, T., Tenhu, H. & Hietala, S. Polyelectrolyte stabilized nanodiamond dispersions. *Diam. Relat. Mater.* **95**, 185–194 (2019).
20. Desai, C., Chen, K. & Mitra, S. Aggregation behavior of nanodiamonds and their functionalized analogs in an aqueous environment. *Environ. Sci. Process. Impacts* **16**, 518–523 (2014).
21. Thomä, S. L. J., Krauss, S. W., Eckardt, M., Chater, P. & Zobel, M. Atomic insight into hydration shells around facetted nanoparticles. *Nat. Commun.* **10**, (2019).

22. Zobel, M. Observing structural reorientations at solvent-nanoparticle interfaces by X-ray diffraction - Putting water in the spotlight. *Acta Cryst.* **A72**, 621–631 (2016).
23. Zobel, M., Neder, R. B. & Kimber, S. A. J. Universal solvent restructuring induced by colloidal nanoparticles. *Science* vol. 347 (2015).
24. Fogarty, A. C. & Laage, D. Water Dynamics in Protein Hydration Shells: The Molecular Origins of the Dynamical Perturbation. *J. Phys. Chem. B* **118**, 53 (2014).
25. Laage, D., Elsaesser, T. & Hynes, J. T. Perspective: Structure and ultrafast dynamics of biomolecular hydration shells. *Struct. Dyn.* **4**, 044018–044018 (2017).
26. Laage, D., Elsaesser, T. & Hynes, J. T. Water Dynamics in the Hydration Shells of Biomolecules. *Chem.Rev.* **117**, 10694–10725 (2017).
27. Lounnas, V., Pettitt, B. M. & Phillips, G. N. A global model of the protein-solvent interface. *Biophys. J.* **66**, 601–614 (1994).
28. Khodadadi, S. & Sokolov, A. P. Atomistic details of protein dynamics and the role of hydration water. *Biochim. Biophys. Acta - Gen. Subj.* **1861**, 3546–3552 (2017).
29. Watanabe, N., Suga, K. & Umakoshi, H. Functional hydration behavior: Interrelation between hydration and molecular properties at lipid membrane interfaces. *Journal of Chemistry* vol. 2019 (2019).
30. Liu, S., Meng, X. Y., Perez-Aguilar, J. M. & Zhou, R. An In Silico study of TiO2 nanoparticles interaction with twenty standard amino acids in aqueous solution. *Sci. Rep.* **6**, (2016).
31. Korobov, M. V, Avramenko, N. V, Bogachev, A. G., Rozhkova, N. N. & Osawa, E. Nanophase of water in nano-diamond gel. *J. Phys. Chem. C* **111**, 7330–7334 (2007).
32. Stehlik, S. *et al.* Water interaction with hydrogenated and oxidized detonation nanodiamonds - Microscopic and spectroscopic analyses. *Diam. Relat. Mater.* **63**, 97–102 (2016).
33. Petit, T. *et al.* Unusual Water Hydrogen Bond Network around Hydrogenated Nanodiamonds. *J. Phys. Chem. C* **121**, 5185–5194 (2017).
34. Petit, T. *et al.* Probing Interfacial Water on Nanodiamonds in Colloidal Dispersion. *J. Phys. Chem. Lett.* **6**, 2909–2912 (2015).
35. Chaux-Jukic, I. *et al.* Revisiting Water Molecule Interactions with Hydrogenated Nanodiamonds: Towards Their Direct Quantification in Aqueous Suspensions. *Am J Nanotechnol Nanomed* **2**, 14–022 (2019).
36. Alizadeh, M., Azar, P. A., Mozaffari, S. A., Karimi-Maleh, H. & Tamaddon, A. M. A DNA Based Biosensor Amplified With ZIF-8/Ionic Liquid Composite for Determination of Mitoxantrone Anticancer Drug: An Experimental/Docking Investigation. *Front. Chem.* **8**, 1–10 (2020).
37. Karimi-Maleh, H. *et al.* Guanine-Based DNA Biosensor Amplified with Pt/SWCNTs Nanocomposite as Analytical Tool for Nanomolar Determination of Daunorubicin as an Anticancer Drug: A Docking/Experimental Investigation. *Ind. Eng. Chem. Res.* **60**, 816–823 (2021).
38. Guo, Y., Li, S., Li, W., Moosa, B. & Khashab, N. M. The Hofmeister effect on nanodiamonds: How addition of ions provides superior drug loading platforms. *Biomater. Sci.* **2**, 84–88 (2014).

39. Zhu, Y. *et al.* Excessive sodium ions delivered into cells by nanodiamonds: Implications for tumor therapy. *Small* **8**, 1771–1779 (2012).
40. Dolenko, T. A. *et al.* Study of adsorption properties of functionalized nanodiamonds in aqueous solutions of metal salts using optical spectroscopy. *J. Alloys Compd.* **586**, (2014).
41. Marcus, Y. Effect of ions on the structure of water: Structure making and breaking. *Chem. Rev.* **109**, 1346–1370 (2009).
42. Hribar, B., Southall, N. T., Vlachy, V. & Dill, K. A. How Ions Affect the Structure of Water. *J Am Chem Soc.* **124**, 12302–12311 (2002).
43. Tobias, D. J. & Hemminger, J. C. Getting Specific about Specific Ion Effects. *Science (80-. ).* **319**, 1197–1198 (2008).
44. Lutter, J. C., Wu, T. Y. & Zhang, Y. Hydration of cations: A key to understanding of specific cation effects on aggregation behaviors of PEO-PPO-PEO triblock copolymers. *J. Phys. Chem. B* **117**, 10132–10141 (2013).
45. Collins, K. D. Sticky Ions in Biological Systems. *PNAS* **92**, 5553–5557 (1995).
46. Marcus, Y. *Ions in water and biophysical implications: From chaos to cosmos*. *Ions in Water and Biophysical Implications: From Chaos to Cosmos* vol. 9789400746473 (Springer Netherlands, 2012).
47. Omta, A. W., Kropman, M. F., Woutersen, S. & Bakker, H. J. Negligible Effect of Ions on the Hydrogen-Bond Structure in Liquid Water. *Science* vol. 301 347–349 (2003).
48. Samoilov, Y. A new approach to the study of hydration of ions in aqueous solutions. *Discuss. Faraday Soc.* **24**, 141–146 (1957).
49. Chen, Y., Okur, H. I., Liang, C. & Roke, S. Orientational ordering of water in extended hydration shells of cations is ion-specific and is correlated directly with viscosity and hydration free energy. *Phys. Chem. Chem. Phys.* **19**, 24678–24688 (2017).
50. Saberi Movahed, F., Cheng, G. C. & Venkatachari, B. S. Atomistic simulation of thermal decomposition of crosslinked and non-crosslinked phenolic resin chains. in *42nd AIAA Thermophysics Conference* (2011).
51. Lo Nostro, P. & Ninham, B. W. Hofmeister phenomena: An update on ion specificity in biology. *Chemical Reviews* vol. 112 2286–2322 (2012).
52. Collins, K. D. Ion hydration: Implications for cellular function, polyelectrolytes, and protein crystallization. *Biophys. Chem.* **119**, 271–281 (2006).
53. Okur, H. I. *et al.* Beyond the Hofmeister Series: Ion-Specific Effects on Proteins and Their Biological Functions. *Journal of Physical Chemistry B* vol. 121 1997–2014 (2017).
54. Jungwirth, P. & Tobias, D. J. Specific ion effects at the air/water interface. *Chemical Reviews* vol. 106 1259–1281 (2006).
55. Aziz, E. F. *et al.* Cation-specific interactions with carboxylate in amino acid and acetate aqueous solutions: X-ray absorption and ab initio calculations. *J. Phys. Chem. B* **112**, 12567–12570 (2008).
56. Rodríguez-Ropero, F. & Fioroni, M. Effect of Na+, Mg2+, and Zn2+ chlorides on the structural and thermodynamic properties of water/n-heptane interfaces. *J. Comput. Chem.* **32**, 1876–1886 (2011).
57. Shenderova, O. *et al.* Surface chemistry and properties of ozone-purified detonation nanodiamonds. *J. Phys. Chem. C* **115**, 9827–9837 (2011).

58. Raty, J. Y. & Galli, G. Optical properties and structure of nanodiamonds. in *Journal of Electroanalytical Chemistry* vol. 584 9–12 (Elsevier, 2005).

59. Manelli, O., Corni, S. & Righi, M. C. Water adsorption on native and hydrogenated diamond (001) surfaces. *J. Phys. Chem. C* **114**, 7045–7053 (2010).

60. Pinto, H. *et al.* First-principles studies of the effect of (001) surface terminations on the electronic properties of the negatively charged nitrogen-vacancy defect in diamond. *Phys. Rev. B - Condens. Matter Mater. Phys.* **86**, (2012).

61. Petit, T. *et al.* Surface transfer doping can mediate both colloidal stability and self-assembly of nanodiamonds. *Nanoscale* **5**, 8958–8962 (2013).

62. Su, L., Krim, J. & Brenner, D. W. Interdependent Roles of Electrostatics and Surface Functionalization on the Adhesion Strengths of Nanodiamonds to Gold in Aqueous Environments Revealed by Molecular Dynamics Simulations. *J. Phys. Chem. Lett.* **9**, 4396–4400 (2018).

63. Plimpton, S. Fast Parallel Algorithms for Short-Range Molecular Dynamics. *J. Comput. Phys.* **117**, 1–19 (1995).

64. Jorgensen, W. L., Maxwell, D. S. & Tirado-Rives, J. Development and Testing of the OPLS All-Atom Force Field on Conformational Energetics and Properties of Organic Liquids. *J. Am. Chem. Soc.* **118**, 11225–11236 (1996).

65. Berendsen, H. J. C., Grigera, J. R. & Straatsma, T. P. The missing term in effective pair potentials. *J. Phys. Chem.* **91**, 6269–6271 (1987).

66. Mark, P. & Nilsson, L. Structure and dynamics of the TIP3P, SPC, and SPC/E water models at 298 K. *J. Phys. Chem. A* **105**, 9954–9960 (2001).

67. Jorgensen, W. L., Chandrasekhar, J., Madura, J. D., Impey, R. W. & Klein, M. L. Comparison of simple potential functions for simulating liquid water. *J. Chem. Phys.* **79**, 926–935 (1983).

68. Joung, I. S. & Cheatham, T. E. Determination of alkali and halide monovalent ion parameters for use in explicitly solvated biomolecular simulations. *J. Phys. Chem. B* **112**, 9020–9041 (2008).

69. Mamatkulov, S., Fyta, M. & Netz, R. R. Force fields for divalent cations based on single-ion and ion-pair properties. *J. Chem. Phys.* **138**, (2013).

70. Hockney, R. W. & Eastwood, J. W. Computer Simulation Using Particles. *Comput. Sci.* (1966).

71. Nikitin, A., Milchevskiy, Y. & Lyubartsev, A. AMBER-II: New Combining Rules and Force Field for Perfluoroalkanes. *J. Phys. Chem. B* **119**, 14563–14573 (2015).

72. Tuckerman, M., Berne, B. J. & Martyna, G. J. Reversible multiple time scale molecular dynamics. *J. Chem. Phys.* **97**, 1990–2001 (1992).

73. Nosé, S. A unified formulation of the constant temperature molecular dynamics methods. *J. Chem. Phys.* **81**, 511–519 (1984).

74. Hoover, W. G. & Holian, B. L. *Kinetic moments method for the canonical ensemble distribution*. *Physics Letters A* vol. 2 (1996).

75. Michaud-Agrawal, N., Denning, E. J., Woolf, T. B. & Beckstein, O. MDAnalysis: A toolkit for the analysis of molecular dynamics simulations. *J. Comput. Chem.* **32**, 2319–2327 (2011).

76. Gowers, R. J. *et al.* MDAnalysis: A Python Package for the Rapid Analysis of Molecular Dynamics Simulations. in *Proc. Of The 15th Python In Science Conf. (SciPy 2016)* (2016).
77. Ding, Y., Hassanali, A. A., Parrinello, M., Designed, M. P. & Per-Formed, A. A. H. Anomalous water diffusion in salt solutions. *PNAS* **111**, (2014).
78. Impey, R. W., Madden, P. A. & McDonald, I. R. Hydration and Mobility of Ions in Solution. *J. Phys. Chem* **87**, 5071–5083 (1983).
79. Debnath, A., Mukherjee, B., Ayappa, K. G., Maiti, P. K. & Lin, S. T. Entropy and dynamics of water in hydration layers of a bilayer. *J. Chem. Phys.* **133**, (2010).
80. Bhide, S. Y. & Berkowitz, M. L. Structure and dynamics of water at the interface with phospholipid bilayers. *J. Chem. Phys.* **123**, (2005).
81. Tiwari, S. P., Rai, N. & Maginn, E. J. Dynamics of actinyl ions in water: A molecular dynamics simulation study. *Phys. Chem. Chem. Phys.* **16**, 8060–8069 (2014).
82. Pizzitutti, F., Marchi, M., Sterpone, F. & Rossky, P. J. How protein surfaces induce anomalous dynamics of hydration water. *J. Phys. Chem. B* **111**, 7584–7590 (2007).
83. Tielrooij, K. J., Garcia-Araez, N., Bonn, M. & Bakker, H. J. Cooperativity in Ion Hydration. *Science* vol. 328 1006–1009 (2010).
84. Tan, H. S., Piletic, I. R. & Fayer, M. D. Orientational dynamics of water confined on a nanometer length scale in reverse micelles. *J. Chem. Phys.* **122**, (2005).
85. Saxena, A. & García, A. E. Multisite ion model in concentrated solutions of divalent cations (MgCl2 and CaCl2): Osmotic pressure calculations. *J. Phys. Chem. B* **119**, 219–227 (2015).
86. Bruni, F., Imberti, S., Mancinelli, R. & Ricci, M. A. Aqueous solutions of divalent chlorides: Ions hydration shell and water structure. *J. Chem. Phys.* **136**, (2012).
87. Mancinelli, R., Botti, A., Bruni, F., Ricci, M. A. & Soper, A. K. Perturbation of water structure due to monovalent ions in solution. *Phys. Chem. Chem. Phys.* **9**, 2959–2967 (2007).
88. Fischer, W. B., Fedorowicz, A. & Koll, A. Structured water around ions - FTIR difference spectroscopy and quantum-mechanical calculations. *Phys. Chem. Chem. Phys.* **3**, 4228–4234 (2001).
89. Larentzos, J. P. & Criscenti, L. J. A molecular dynamics study of alkaline earth metal-chloride complexation in aqueous solution. *J. Phys. Chem. B* **112**, 14243–14250 (2008).
90. Liu, C., Min, F., Liu, L. & Chen, J. Hydration properties of alkali and alkaline earth metal ions in aqueous solution: A molecular dynamics study. *Chem. Phys. Lett.* **727**, 31–37 (2019).
91. Hofmann, A. E., Bourg, I. C. & DePaolo, D. J. Ion desolvation as a mechanism for kinetic isotope fractionation in aqueous systems. *Proc. Natl. Acad. Sci. U. S. A.* **109**, 18689–18694 (2012).
92. Mancinelli, R., Botti, A., Bruni, F., Ricci, M. A. & Soper, A. K. Hydration of sodium, potassium, and chloride ions in solution and the concept of structure maker/breaker. *J. Phys. Chem. B* **111**, 13570–13577 (2007).
93. Ohtaki, H. & Radnai, T. Structure and Dynamics of Hydrated Ions. *Chem. Rev* **93**, 1157–1204 (1993).
94. Lee, Y., Thirumalai, D. & Hyeon, C. Ultrasensitivity of Water Exchange Kinetics to the Size of Metal Ion. *J. Am. Chem. Soc.* **139**, 12334–12337 (2017).

95. Koskamp, J. A. *et al.* Reconsidering Calcium Dehydration as the Rate-Determining Step in Calcium Mineral Growth. *J. Phys. Chem. C* **123**, 26895–26903 (2019).
96. Helm, L. & Merbach, A. E. Water exchange on metal ions: experiments and simulations. *Coord. Chem. Rev.* **187**, 151–181 (1999).
97. Bleuzen, A., Pittet, P.-A., Helm, L. & Merbach, A. E. Water Exchange on in Aqueous Magnesium(II) Solution : a Variable Temperature and Pressure 17O NMR Study. *Magn. Reson. Chem.* **35**, 765–773 (1997).
98. Zajforoushan Moghaddam, S. & Thormann, E. The Hofmeister series: Specific ion effects in aqueous polymer solutions. *J. Colloid Interface Sci.* **555**, 615–635 (2019).
99. Wang, L. Y., Zhang, Y. H. & Zhao, L. J. Raman spectroscopic studies on single supersaturated droplets of sodium and magnesium acetate. *J. Phys. Chem. A* **109**, 609–614 (2005).
100. Pastorczak, M., Van Der Post, S. T. & Bakker, H. J. Cooperative hydration of carboxylate groups with alkali cations. *Phys. Chem. Chem. Phys.* **15**, 17767–17770 (2013).
101. Stirnemann, G., Wernersson, E., Jungwirth, P. & Laage, D. Mechanisms of Acceleration and Retardation of Water Dynamics by Ions. *J. Am. Chem. Soc.* **135**, 11824–11831 (2013).
102. Van Der Vegt, N. F. A. *et al.* Water-Mediated Ion Pairing: Occurrence and Relevance. *Chem. Rev.* **116**, 7626–7641 (2016).
103. Laage, D., Stirnemann, G., Sterpone, F., Rey, R. & Hynes, J. T. Reorientation and Allied Dynamics in Water and Aqueous Solutions. *Annu. Rev. Phys. Chem.* **62**, 395–416 (2011).
104. Laage, D. & Hynes, J. T. Reorientional dynamics of water molecules in anionic hydration shells. *PNAS* **104**, 11167–11172 (2007).
105. Bock, C. W., Markham, G. D., Katz, A. K. & Glusker, J. P. The arrangement of first- and second-shell water molecules around metal ions: Effects of charge and size. *Theor. Chem. Acc.* **115**, 100–112 (2006).
106. Catalano, J. G. Weak interfacial water ordering on isostructural hematite and corundum (001) surfaces. *Geochim. Cosmochim. Acta* **75**, 2062–2071 (2011).
107. Spagnoli, D., Gilbert, B., Waychunas, G. A. & Banfield, J. F. Prediction of the effects of size and morphology on the structure of water around hematite nanoparticles. *Geochim. Cosmochim. Acta* **73**, 4023–4033 (2009).
108. Kerisit, S., Cooke, D. J., Spagnoli, D. & Parker, S. C. Molecular dynamics simulations of the interactions between water and inorganic solids. *J. Mater. Chem.* **15**, 1454–1462 (2005).
109. Vila Verde, A. & Lipowsky, R. Cooperative slowdown of water rotation near densely charged ions is intense but short-ranged. *J. Phys. Chem. B* **117**, 10556–10566 (2013).
110. Garcia, A. E. & Stiller, L. Computation of the Mean Residence Time of Water in the Hydration Shells of Biomolecules. *J. Comput. Chem.* **14**, 1396–1406 (1993).
111. Batsanov, S. S. & Batsanov, A. S. Effect of diamond on structure and properties of confined water. *Chem. Phys. Lett.* **651**, 8–12 (2016).
112. Nickolov, Z., Ivanov, I., Georgiev, G. & Stoilova, D. Raman study of complexation in aqueous solutions of magnesium acetate. *J. Mol. Struct.* **377**, 13–17 (1996).
113. Semmler, J., Irish, D. E. & Ozeki, T. Vibrational spectral studies of solutions at elevated temperatures and pressures. 12. Magnesium acetate. *Geochim. Cosmochim. Acta* **54**, 947–954 (1990).

114. Wahab, A. *et al.* Ultrasonic velocities, densities, viscosities, electrical conductivities, raman spectra, and molecular dynamics simulations of aqueous solutions of Mg(OAc)2 and Mg(NO3)2: Hofmeister effects and ion pair formation. *J. Phys. Chem. B* **109**, 24108–24120 (2005).
115. Collins, K. D. Charge Density-Dependent Strength of Hydration and Biological Structure. *Biophys. J.* **72**, 65–76 (1997).
116. Collins, K. D. The behavior of ions in water is controlled by their water affinity. *Q. Rev. Biophys.* **52**, e11 (2019).
117. Mehandzhiyski, A. Y., Riccardi, E., Van Erp, T. S., Trinh, T. T. & Grimes, B. A. Ab Initio Molecular Dynamics Study on the Interactions between Carboxylate Ions and Metal Ions in Water. *J. Phys. Chem. B* **119**, 10710–10719 (2015).

# Supplementary Information

In this section, the average values of the residence time of water, whose plots were shown in Section 3.3 of the paper, in three hydration layers of various DNDs solvated in different salt solutions are tabulated. The average values are obtained from five independent MD trajectories. Furthermore, the 95% confidence interval (CI) for each average value is also calculated using the bootstrap percentile method[1]. The bootstrap distribution was generated by resampling the sample data 10,000 times with replacement.

**Table 0S.1.** The average values of the residence time of water, $\tau_{res}^{(1)}$, $\tau_{res}^{(2)}$, and $\tau_{res}^{(3)}$ that correspond to, respectively, the first, the second, and the third hydration layers of DND–H. Numbers in parentheses denote the bootstrapped CI.

| DND's Charge | Dissolved Salt | $\tau_{res}^{(1)}$ / ps | $\tau_{res}^{(2)}$ / ps | $\tau_{res}^{(3)}$ / ps |
|---|---|---|---|---|
| q = 0 | KCl | 8.7696 (8.5960, 8.9140) | 4.3709 (4.3068, 4.4349) | 2.5937 (2.5751, 2.6242) |
| | NaCl | 8.6404 (8.5138, 8.7172) | 4.3713 (4.3256, 4.4242) | 2.3184 (2.3005, 2.3461) |
| | CaCl$_2$ | 8.9213 (8.7988, 9.0787) | 4.0952 (4.0369, 4.1510) | 2.5923 (2.5697, 2.6199) |
| | MgCl$_2$ | 8.9794 (8.8511, 9.1110) | 3.8813 (3.8386, 3.9263) | 2.6496 (2.6012, 2.6976) |
| q = +28 | KCl | 11.6849 (11.3893, 11.9547) | 4.5726 (4.5185, 4.6303) | 2.8968 (2.8706, 2.9175) |
| | NaCl | 11.2576 (11.0724, 11.4112) | 4.7121 (4.6189, 4.8245) | 2.7635 (2.7355, 2.7849) |
| | CaCl$_2$ | 11.4406 (11.3277, 11.5607) | 4.9959 (4.9546, 5.0408) | 2.5649 (2.5349, 2.5977) |
| | MgCl$_2$ | 11.309 (10.9504, 11.5034) | 4.7721 (4.6851, 4.8331) | 3.194 (3.1591, 3.2224) |
| q = +56 | KCl | 11.5568 (11.3113, 11.7732) | 5.442 (5.3403, 5.5962) | 3.2729 (3.2471, 3.3063) |
| | NaCl | 12.6958 (12.4971, 13.0002) | 5.1584 (5.1308, 5.1886) | 3.4409 (3.3863, 3.4955) |
| | CaCl$_2$ | 12.2903 (12.0663, 12.5287) | 5.0665 (4.9431, 5.1608) | 3.3431 (3.2788, 3.4026) |
| | MgCl$_2$ | 12.6374 (12.4446, 12.8126) | 5.3834 (5.2563, 5.5686) | 3.2239 (3.1770, 3.2931) |
| q = +84 | KCl | 13.9161 (13.8005, 14.0281) | 6.5041 (6.4167, 6.6123) | 3.3132 (3.2864, 3.3434) |
| | NaCl | 13.3079 (13.0472, 13.5090) | 6.4493 (6.3928, 6.5138) | 3.2572 (3.1942, 3.3203) |
| | CaCl$_2$ | 13.6533 (13.4218, 13.9765) | 5.8335 (5.7669, 5.9001) | 3.6162 (3.5559, 3.6747) |
| | MgCl$_2$ | 12.0859 (11.8199, 12.4444) | 6.1387 (6.0788, 6.1897) | 3.4676 (3.4308, 3.5043) |

**Table 0.2.** Same as Table 0S.1, but for the case of DND–NH$_2$.

| DND's Charge | Dissolved Salt | $\tau_{res}^{(1)}$ / ps | $\tau_{res}^{(2)}$ / ps | $\tau_{res}^{(3)}$ / ps |
|---|---|---|---|---|
| q = 0 | KCl | 19.8886 (19.4640, 20.3830) | 3.9361 (3.8815, 3.9933) | 3.5662 (3.5251, 3.6073) |
| | NaCl | 19.5192 (19.0366, 20.1512) | 4.1178 (4.0819, 4.1570) | 3.8353 (3.8035, 3.8681) |
| | CaCl$_2$ | 19.0089 (18.7191, 19.2187) | 4.3169 (4.2690, 4.3543) | 3.4211 (3.3574, 3.4692) |
| | MgCl$_2$ | 20.7124 (19.6620, 21.6841) | 4.1171 (4.0066, 4.2430) | 3.2454 (3.1740, 3.3355) |
| q = +28 | KCl | 19.4758 (19.2335, 19.7300) | 5.3662 (5.3128, 5.4064) | 2.7967 (2.7705, 2.8361) |
| | NaCl | 19.4442 (18.8315, 20.1200) | 5.0541 (4.9851, 5.1231) | 3.8464 (3.7955, 3.9014) |
| | CaCl$_2$ | 20.3721 (19.8922, 20.7787) | 5.0627 (5.0159, 5.1268) | 3.3804 (3.3415, 3.4158) |
| | MgCl$_2$ | 19.6553 (19.0637, 20.1341) | 5.2496 (5.1763, 5.3445) | 3.1219 (3.0896, 3.1492) |
| q = +56 | KCl | 34.5115 (33.5170, 35.4007) | 5.9188 (5.8091, 6.0438) | 2.8769 (2.8533, 2.9031) |
| | NaCl | 33.2447 (32.0979, 34.4506) | 5.4421 (5.3435, 5.5613) | 3.1159 (3.0719, 3.1662) |
| | CaCl$_2$ | 34.276 (33.0005, 35.6497) | 5.9863 (5.8654, 6.1173) | 3.4104 (3.3265, 3.4815) |
| | MgCl$_2$ | 33.9984 (32.7877, 35.7208) | 5.5821 (5.4438, 5.7227) | 3.7502 (3.7375, 3.7660) |
| q = +84 | KCl | 65.7183 (58.9919, 70.9438) | 9.9659 (9.8466, 10.1265) | 3.3775 (3.3480, 3.4211) |
| | NaCl | 66.9624 (61.6642, 72.8107) | 9.5798 (9.4816, 9.7022) | 3.795 (3.7449, 3.8336) |
| | CaCl$_2$ | 70.221 (67.1431, 73.0547) | 8.9746 (8.8735, 9.0869) | 4.1496 (4.1184, 4.1829) |
| | MgCl$_2$ | 75.8178 (72.3780, 79.9321) | 9.2409 (9.1403, 9.3701) | 3.7241 (3.7003, 3.7503) |

**Table 0.3.** Same as Table 0S.1, but for the case of DND–COOH.

| DND's Charge | Dissolved Salt | $\tau_{res}^{(1)}$ / ps | $\tau_{res}^{(2)}$ / ps | $\tau_{res}^{(3)}$ / ps |
|---|---|---|---|---|
| q = 0 | KCl | 22.4753 (21.7155, 23.0924) | 6.4011 (6.3443, 6.4768) | 2.9629 (2.9452, 2.9805) |
| | NaCl | 22.5284 (22.1581, 23.0570) | 6.3382 (6.3006, 6.3884) | 3.5139 (3.4873, 3.5386) |
|  | CaCl$_2$ | 24.7294 (23.5696, 25.6512) | 6.6171 (6.5422, 6.6912) | 3.8564 (3.7855, 3.9249) |
|  | MgCl$_2$ | 23.2814 (22.7015, 24.0326) | 6.6087 (6.5736, 6.6409) | 3.5202 (3.4479, 3.6096) |
| q = –28 | KCl | 22.2157 (21.3730, 23.2283) | 6.7708 (6.6802, 6.8982) | 4.0625 (4.0386, 4.0854) |
|  | NaCl | 25.8 (24.6611, 26.7899) | 6.3876 (6.2737, 6.5015) | 4.088 (4.0288, 4.1434) |
|  | CaCl$_2$ | 29.8023 (29.2486, 30.3588) | 7.4163 (7.3188, 7.5016) | 4.0446 (4.0278, 4.0614) |
|  | MgCl$_2$ | 30.2519 (29.2700, 31.1416) | 9.3696 (9.1769, 9.5622) | 3.2749 (3.1985, 3.3344) |
| q = –56 | KCl | 26.5493 (25.5549, 27.8003) | 7.2768 (7.1896, 7.3567) | 4.0805 (4.0557, 4.1064) |
|  | NaCl | 27.9387 (26.9087, 28.7888) | 6.386 (6.2810, 6.5107) | 3.7467 (3.7124, 3.7797) |
|  | CaCl$_2$ | 33.6438 (32.9433, 34.4457) | 8.1101 (7.7494, 8.4095) | 4.3028 (4.2337, 4.3571) |
|  | MgCl$_2$ | 33.4188 (32.0399, 35.0134) | 11.6672 (11.5100, 11.7895) | 3.4274 (3.3716, 3.4832) |
| q = –84 | KCl | 27.9886 (26.5494, 29.9288) | 8.9632 (8.8758, 9.0459) | 3.9673 (3.9390, 3.9942) |
|  | NaCl | 34.8591 (32.6337, 36.6374) | 7.2222 (7.0323, 7.4388) | 3.1382 (3.0916, 3.1961) |
|  | CaCl$_2$ | 53.261 (50.8361, 56.0259) | 10.5079 (10.0197, 10.9478) | 4.005 (3.9661, 4.0488) |
|  | MgCl$_2$ | 51.2375 (47.8200, 55.1412) | 14.0457 (13.6343, 14.4077) | 4.0671 (4.0148, 4.1203) |

**Table 0.4.** Same as Table 0S.1, but for the case of DND–OH.

| DND's Charge | Dissolved Salt | $\tau_{res}^{(1)}$ / ps | $\tau_{res}^{(2)}$ / ps | $\tau_{res}^{(3)}$ / ps |
|---|---|---|---|---|
| q = 0 | KCl | 14.5962 (14.4948, 14.6976) | 5.0529 (4.9973, 5.1318) | 3.4329 (3.3877, 3.4768) |
|  | NaCl | 13.4393 (13.1881, 13.6539) | 5.2695 (5.1494, 5.4089) | 2.9836 (2.9467, 3.0252) |
|  | CaCl$_2$ | 13.7912 (13.4245, 14.1818) | 5.4795 (5.2767, 5.6842) | 3.6808 (3.6367, 3.7296) |
|  | MgCl$_2$ | 15.248 (14.9581, 15.5379) | 4.8799 (4.8133, 4.9640) | 3.1115 (3.0121, 3.1981) |

**References**

1.      Efron, B. & Tibshirani, R. An Introduction to the Bootstrap. (Chapman and Hall, 1993).